\documentclass[10pt,twocolumn,twoside] {IEEEtran}

\usepackage{pifont}
\usepackage{stmaryrd}
\usepackage{mathrsfs}
\usepackage{amsmath}
\usepackage{amssymb}
\usepackage{multirow}
\usepackage{cite}
\usepackage[dvips]{graphicx}
\usepackage[ruled]{algorithm2e}
\usepackage{array}

\newcommand{\be}{\begin{equation}}
\newcommand{\ee}{\end{equation}}
\newcommand{\bp}{\mbox{\boldmath{$p $}}}
\newcommand{\bv}{\mbox{\boldmath{$v $}}}
\newcommand{\br}{\mbox{\boldmath{$r $}}}

\newcommand{\btheta}{{\mbox{\boldmath{$\theta$}}}}

\begin{document}
\title{Optimal Deployment of Multistatic Radar System Using Multi-Objective Particle Swarm Optimization}

\author{Yichuan Yang, Tianxian Zhang, Wei Yi, Lingjiang Kong, Xiaolong Li, Bing Wang, Xiaobo Yang
\thanks{Y. Yang, T. Zhang, W. Yi, L. Kong (Corresponding author), X. Li, B. Wang and X. Yang
are with University of Electronic Science and Technology of China, School of
Electronic Engineering, Chengdu, Sichuan, P.R. China, 611731. Fax:
+86-028-61830064, Tel: +86-028-61830768, E-mail: lingjiang.kong@gmail.com.}}

\maketitle

\begin{abstract}
We consider an optimization deployment problem of multistatic radar system (MSRS). Through the antenna placing and the transmitted power allocating, we optimally deploy the MSRS for two goals: 1) the first one is to improve the coverage ratio of surveillance region; 2) the second goal is to get a even distribution of signal energy in surveillance region. In two typical working modes of MSRS, we formulate the optimization problem by introducing two objective functions according to the two mentioned goals, respectively. Addressing on two main challenges of applying multi-objective particle swarm optimization (MOPSO) in solving the proposed optimization problem, we propose a deployment algorithm based on multi-objective particle swarm optimization with non-dominated relative crowding distance (MOPSO-NRCD). For the challenge of value difference, we propose a novel selection method with a non-dominated relative crowding distance. For the challenge of particle allocation, a multi-swarm structure of MOPSO is also introduced. Finally, simulation results are given out to prove the advantages and validity of the proposed deployment algorithm. It is shown that with same number of employed particles, the proposed MOPSO-NRCD algorithm can achieve better optimization performance than that of traditional multi-objective particle swarm optimization with crowding distance (MOPSO-CD).
\end{abstract}


\maketitle

\section{Introduction}
Taking advantages of higher amount of information available, multistatic radar system (MSRS) has been proved with ability to improve performance in many aspects (e.g., detection and estimation) \cite{Griffiths,Baker,Haimovich,QHe,Wang,AMaio,Godrich}. Meanwhile, such radar system has potential capability to be effective measures to counter the threats of stealth targets and anti-radar missiles in military applications \cite{Neng}. Comparing with the conventional radar with a single transmit antenna, MSRS provides more degrees of freedom, which support flexible management modes, lead to further improved on many aspect of performance \cite{Haimovich}. In fact, the management of MSRS, which can be divides into many modes, not only improves but also affects the performance of MSRS to a larger extend. Therefore, the management of MSRS should be optimized in order to get desire performances.

Within all the management modes, the deployment of MSRS, which involved with massive parameter settings and adjustment (e.g., placing the dispersed antennas and allocating transmitted power), is of great importance to be a basis that affect the performance of MSRS to a large extent \cite{He,Godrich2011,Chen,Chavali,Radmard,chuan}. Specifically, the antenna optimal placement \cite{He,Chen,Radmard,chuan} and power optimal allocation \cite{Godrich2011,Chavali,Radmard}, which are two main aspects of MSRS deployment, have been proved with the ability to improve the localization \cite{He,Godrich2011,Chen,Chavali} and detection \cite{Radmard,chuan} performances of MSRS. For the purpose of providing a proper basis for MSRS to fully explore its potential (e.g., better performance and more flexibility), the optimal deployment of MSRS should be ascertained. As we know, the establishment of the optimal deployment can be translated to an optimization problem with two key elements such as \textbf{the optimization model} and \textbf{the optimization algorithm}. Many investigations have been done on these two elements.

\subsection{Overview}
\textbf{For the optimization model:} The optimization model is important as it determines the deployment purpose and variables that can be adjusted. Specifically, by employing different performance metrics as the objectives, different optimal deployments of MSRS can be obtained. Some published works focus on the localization performance metrics \cite{He,Godrich2011,Chen,Chavali}. The derived Cram\'{e}r-Rao bound (CRB) on velocity estimation of target \cite{He} and the target location estimation accuracy of an ultrawideband multiple-input-multiple-output (MIMO) noise radar system \cite{Chen} are introduced as objectives to explore the optimal antenna placement of MSRS. In \cite{Godrich2011} and \cite{Chavali}, the optimal power allocations are obtained by considering two objectives, i.e., the CRB on target position estimation and the posterior CRB on target state estimation. Besides, detection performance metrics are also been taken into account, where a weighted average detection performance and a coverage ratio are respectively adopted as objectives in \cite{Radmard,chuan}. For the investigating of MSRS deployment problem, the above mentioned works just focus on one objective. However, radar systems usually need to simultaneously adapt to different situations, wherein different performance metrics are mainly concerned. Therefore, the deployment of MSRS need to simultaneously take more than one objective into account.

\textbf{For the optimization algorithm:} The optimization algorithm is another important issue since it determines the optimization performance. In \cite{He}, the optimal antenna placement is obtained from an derived analytical expression of CRB on target velocity estimation. For the purpose of achieving performance better than a predefined threshold by optimally allocating the power, a relaxation and domain decomposition method is introduced in \cite{Godrich2011}. In addition, a geometrical and numerical method is employed to improve the estimation accuracy of target location in \cite{Chen}.

The above mentioned works only consider the situation that a target locates in one specific resolution cell. In real application, instead of focusing on one single resolution cell, it is more practical to consider the MSRS deployment problem for a region which contains multiple resolution cells. However, by considering a region with multiple resolution cells and taking some performance metrics as objectives, the deployment problems of MSRS are normally of great computational complexity (i.e., huge computational load). Considering the detection performance of MSRS in a region, a sequentially exhaustive enumeration (SEE) method is introduced in \cite{Radmard}. While the SEE is still suffered the challenge of a huge computational load, especially when the concerned region contains many resolution cells and there are many variables to be adjusted (i.e., many antenna to be placed). To address this problem, we propose a particle swarm optimization (PSO) based placement algorithm of MSRS in our previous work \cite{chuan}. The computational load can be significantly reduced.

However, the above works pay no attention to a more general situation, wherein more than one objective are simultaneously considered. Therefore, the optimization problem becomes a multi-objective problem (MOP), which is normally more complex than single-objective optimization problem \cite{Ke,Jiang,Shim}. Recently, to expand the PSO algorithm in solving MOP, i.e., multi-objective particle swarm optimization (MOPSO), has got an increasing interest in both theoretical research \cite{Coello2002,Parsopoulos,Coello2004,Raquel,Reyes} and real application \cite{Goudos,Chamaani,Bing,Pradhan}. For the researches on real applications, some works take multiple antenna metrics (e.g., the side lobe levels, antenna gains, the reflection coefficient over the operating frequency band, and so on) as objectives in \cite{Goudos,Chamaani}. In \cite{Pradhan}, the coverage and lifetime are taken into consideration to study the deployment of sensor networks. Nevertheless, none of these works consider a practical and common case, in which the concerned objectives are significant different to each other (e.g., values of multiple objective functions).

\subsection{Motivation}
The motivations of the current paper are respectively focus on the two above aspects:

1) In real radar applications, radar systems usually need to adapt to different situations in a working environment of continuous changes, wherein different tasks and signal processings are required. Thus, to enhance the utilization efficiency of resources and flexibility of system, the deployment of MSRS should provide a basis contains improvement of objectives characterized by different performance metrics, which can evaluate or affect the performance of different radar tasks.

2) Considering the performances of MSRS in a region, the objectives of the optimization deployment problem may be significantly different to each other. Then, the selection method of global best solution, which plays a key role to determines the optimization performance of MOPSO to a lager extend, becomes challenging since it is easily to be effected by the concerned objectives. Based on the concept of nearest neighbor density estimation \cite{Reyes}, a typical and widely used method with crowding distance is introduced to select the global best solution in order to guide the search of MOPSO \cite{Raquel}. While obtained by directly summing all components (i.e., sub-crowding-distances in the dimensionality of all objectives), the crowding distance can be called as absolute crowding distance, substantially. However, direct sum of all components is mainly effected by the components whose values are much larger than the rests (e.g., the differences between values of multiple objective functions could be several orders of magnitude). Therefore, the absolute crowding distance can not adequately show the true crowding information in some practical situations, wherein the values of multiple objective functions as well as the sub-crowding-distances of objectives are significantly different. Such problem is very common in real application including the deployment of MSRS, while there is no effective method to solve it.

\subsection{Original Contributions}
The original contributions of this paper are as follows.

1) We propose a general optimization deployment model, in which different adjustable variables and objectives can be taken into account for similar systems (e.g., sensor networks) with different purposes. Then, for the purpose of improving the surveillance performance of MSRS equipped with different detection methods, we introduce a coverage ratio and a lowest ratio of total signal energy to noise power as objective functions to formulate a specific optimization problem, wherein the antenna positions and transmitted power are considered as adjustable variables for deployment of MSRS.

2) Since the values of multiple objective functions are commonly of significant difference in real applications, we propose a non-dominated\footnote{Taking maximization problem as example, given two vectors \(\boldsymbol{\alpha},\boldsymbol{\beta}\), if \(\boldsymbol{\alpha} \geqslant \boldsymbol{\beta}\) for \(\alpha_k \geqslant \beta_k, k=1,2,\dots,K\). Then, we say that \(\boldsymbol{\alpha}\) dominates \(\boldsymbol{\beta}\) (denotes by \(\boldsymbol{\alpha}\prec \boldsymbol{\beta}\)) if \(\boldsymbol{\alpha} \geqslant \boldsymbol{\beta}\) and \(\boldsymbol{\alpha} \neq \boldsymbol{\beta}\) \cite{Reyes}.} relative crowding distance based method to properly select global best solution. Specifically, we firstly select candidates from the non-dominated solutions. The sub-crowding-distances of these selected candidates in dimensionality of all concerned objectives are non-dominated. Then, we calculate the relative crowding distance by normalizing and summing all the sub-crowding-distances for each candidate solution. Finally, global best solutions will be selected out from the candidates according to their relative crowding distances.

\subsection{Organization}
The rest of this paper is organized as follows. In Section II, we formulate the optimization problem for MSRS deployment by choosing two typical working modes and proposing two objective functions according to the concerned objectives. Then, focusing on the challenge that the values of multiple objective functions are of great difference, we propose a novel MOPSO algorithm based on non-dominated relative crowding distance in Section III. The presented method is demonstrated by using some numerical examples in Section IV. Finally, the conclusion is given in Section V.

\section{Problem Formulation}\label{sect:problem}
\begin{figure}[t]
\centering
\includegraphics[width=2.8in]{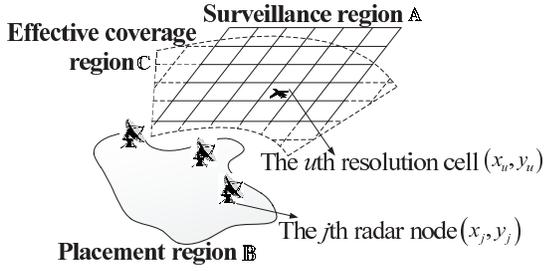}
\caption{Sketch of surveillance region, placement region and effective coverage region.}
\label{fig: region}
\end{figure}

Though with difficulty due to the setting and adjustments of massive variables (e.g., placing antennas and allocating transmitting power), the deployment is of a great importance to be a foundation of MSRS and a determination for the performances of MSRS to a large extent. For the purpose of fully exploring the potential and achieving the desirable performances, the optimal deployment should be established. Such establishment could be usually formulated as an optimization problem in a general form:
\begin{align}\label{Problem Generally formulation}
\hat{\mathbf{\Theta}} = \text{arg} \max \limits_{\mathbf{\Theta}} \mathbf{G} (\mathbf{\Theta})
:=[g_1(\mathbf{\Theta}),g_2(\mathbf{\Theta}),\dots,g_K(\mathbf{\Theta})],
\end{align}
where \(g_k(\mathbf{\Theta})\) (\(k=1,2,\dots,K\)) denotes the \(k\)th objective characterised by some performance metrics, \(K\) is the number of concerned objectives. \(\mathbf{\Theta}\) is a vector that contains the adjustable variables for MSRS deployment. We refer to \(\mathbf{\Theta}\) as a deployment vector (\textbf{DV}). We also refer to \(\mathbf{G} (\mathbf{\Theta})\) as a objective vector (\textbf{OV}) which contains all the concerned objectives. In practical applications, radar systems usually need to face the requirement of different tasks whose performances are evaluated and effected by different metrics. To fully consider all the performance metrics may make the problem too complex to be solved. Moreover, in real radar application, instead of taking all missions into account at the same period, radar systems usually just need to focus on the major missions and ignore the insignificant missions. Thus, in this paper, we focus on the radar detection performance to study the MSRS deployment problem. Meanwhile, our optimization model and algorithms can be easily expanded to other radar applications or missions.
\subsection{Selection of Objectives and Optimization Parameters}\label{sect:obj1}
Generally, a common duty of radar systems in both civil and military applications is monitoring a surveillance region with targets of interest or potential threats. To fulfill such duty, radar systems should be capable to reliably detect targets in arbitrary resolution cell of a surveillance region. As we all know, the effectiveness of target detection can be usually regarded as that the detection probability (\(P_d\)) of a target in a resolution cell is higher than a preset threshold (\(P_{dt}\)) \cite{Merrill}. Then, the radar surveillance performance is determined by the effectiveness of target detection for all the resolution cells within the surveillance region. Similarly, surveillance performance of MSRS is also determined by target detection. Thus, we need to find proper objectives to optimally deploy the MSRS. Similar to our previous work \cite{chuan}, we select the coverage ratio (\(\mathcal{C_R}\)) as one objective of the optimal MSRS deployment problem.

To better understand, the introduction of \(\mathcal{C_R}\) begins with the definition of effective coverage region (ECR). If the \(P_d\) of a target in some resolution cells is higher than \(P_{dt}\) (i.e., \(P_d \geqslant P_{dt}\)), we call these cells as effectively covered resolution cell. The ECR is constructed by all effectively covered resolution cells. Under a commonly setting of MSRS, resolution cells of MSRS are usually approximated as square grids \cite{Haimovich}. Let \(\mathbb{A}\) and \(\mathbb{C}\) denote the surveillance region and the effective coverage region, respectively (see Fig.\ref{fig: region}). Then, we define the coverage ratio of surveillance region (\(\mathcal{C}_\mathcal{R}\)) as:
\begin{align}\label{eq: CR}
\ \mathcal{C}_\mathcal{R} = \frac {\textrm{area}(\mathbb{C} \cap \mathbb{A})}{\textrm{area}( \mathbb{A})}.
\end{align}
Obviously, a bigger \(\mathcal{C}_\mathcal{R}\) represents a better reliable detection performance and the radar system will effectively monitor the whole surveillance region (i.e., \(\mathbb{A}\)) when the \(\mathcal{C}_\mathcal{R}\) reaches \(1\) (i.e., \(\mathbb{C} \cap \mathbb{A}=\mathbb{A}\)). The goal of \(\mathcal{C_R}=1\) may be not difficult to achieve for a strong target. However, such goal is extremely hard for a small target. It is because that the signal energy to noise power ratio of the small target is usually very low, which decreases its detection probability as well as the \(\mathcal{C_R}\). To address this problem, some detection algorithms have been proposed by increasing the complexity of signal processing, such as track-before-detect (TBD) \cite{Tonissen,WYi} and adaptive waveform design technologies \cite{Calderbank,Tian}. However, due to lacking real-time capabilities \cite{WYi}, small application area \cite{Tian} and huge computation complexity \cite{Tonissen,Calderbank,Tian}, these detection algorithms is not suitable for real time monitoring of a surveillance region. Recently, we have proposed a method, which can significantly increase the \(\mathcal{C_R}\) by optimally placing the antennas of MSRS \cite{chuan}. Nevertheless, as just only focusing on \(\mathcal{C_R}\), the signal energy will be unevenly distributed in the surveillance region. Thus, in some parts of the surveillance region, the signal energy to noise power ratio could be reduced to a excessively low level, which would degrade the detection performance sharply. Therefore, some radar coverage gaps that can not be effectively monitored will occur. Such phenomenon is usually unacceptable to the surveillance requirement of a radar system.

To avoid this problem, we propose a solution contains two deployment goals to improve the surveillance performance of radar system: 1) The first goal is to enhance the \(\mathcal{C_R}\) of radar system; 2) for the purpose of getting a even distribution of signal energy in surveillance region, which can provide a proper basis to facilitate the usage of the complex methods to achieve desirable performance, the second goal is to roundly improve the ratio of total signal energy to noise power (RTSN). To this end, based on worst-case performance optimization, we choose to enhance the RTSN of the received returning signal whose RTSN is the lowest among others. Assume there are \(U\) resolution cells in surveillance region, the second objective can be expressed as:
\begin{align}\label{eq: LS}
\mathcal{L}_\mathcal{R}= \min \limits_{u=1,2,\dots,U}\chi_u,
\end{align}
where \(\mathcal{L}_\mathcal{R}\) denotes the lowest RTSN of all resolution cells in surveillance region, \(\chi_u\) denotes the RTSN of the returning signals from the \(u\)th resolution cell.

After the introduction of the goals, their corresponding objective functions as well as the calculation methods are needed to build the problem, which will be introduced in the following subsections.

\subsection{Introductions of Working Modes of MSRS}\label{sect:obj2}
Due to the existing of various adjustable parameters, the MSRS is a system of high degree of freedom and it has many different working modes, under which the calculation methods of one objective could be different. According to the reception mechanism of return signals, MSRS roughly have two typical working modes and we briefly introduce them here.

\subsubsection{\textbf{Cooperative Mode}}\label{sect:c model}
In this mode, the signals transmitted by transmitters of different radar nodes are orthogonal to each other, each receiver of  radar nodes can receive and separate the returning signals reflected by targets. All the received returning signals will be jointly processed. Taking advantages of the cooperation of dispersed radar nodes to improve performance in many aspects, this working mode has received a significant attention in recent years\cite{Haimovich,QHe,Wang,Godrich,He,Godrich2011}. However, in real applications of such mode, the control and synchronize of the widely dispersed radar nodes are difficult (e.g., the problem of pulse chasing will occur when controlling all the radar nodes to illuminate a region simultaneously \cite{Chernyak}). Meanwhile, to satisfied the requirement of cooperation (e.g., communication requirement), system load of this working mode would be big.

\subsubsection{\textbf{Non-cooperative Mode}}\label{sect:nc model}
In this mode, each radar node of MSRS works independently without cooperation with others and it can only receive and process the signal transmitted by itself. Though without performance improvement of cooperation, such working mode does not have to settle the challenge of controlling and synchronization of dispersed radar nodes. Meanwhile, the system resources (e.g., computation and communication resources) consumption of this mode is much less than that of cooperative mode, which facilitates its application in real application. Thus, non-cooperative mode is an effective supplement to cooperative mode and we need to take it into account when considering the optimal deployment of MSRS.

\subsection{Introduction of Objective Functions}\label{sect:method}
After the introduction of deployment goals (\(\mathcal{C_R}\) and \(\mathcal{L_R}\)), we still need their objective functions to solve the problem. In this subsection, under the two typical working mode, we propose the corresponding objective functions as well as their calculation methods for the two deployment goals.

\subsubsection{\textbf{Objective function of} \(\mathcal{C_R}\)}\label{sect:cov}
Since the shape of ECR of MSRS is normally irregular, we have to analyse the \(P_d\) of a target for each resolution cell within a surveillance region in order to get \(\mathcal{C_R}\). Meanwhile, the surveillance region often contains massive resolution cells and the calculation of \(P_d\) of MSRS is difficult in both theoretical research (e.g., complex integral) \cite{AMaio} and real application (e.g., phase and amplitude information of different transmit-receive (T-R) pairs are inconsistent) \cite{Haimovich}. Therefore, the calculation of \(\mathcal{C_R}\) is usually of great computational complexity.

To reduce the computational complexity, the square-law detector (as a commonly used non-coherent detector) is employed for MSRS. Specifically, square-law detector firstly sums the square of samples of the received returning signals, then compares it with a fixed threshold to detect the existence of targets \cite{Merrill}. Such detector has two major advantages: 1) We can acquire a simple \(P_d\) calculation method of this detector under some reasonable assumptions. 2) Without considering the phase information, which avoid the difficulty of phase aligning of signals of different T-R pairs. Therefore, the square-law detector is easy to implement for real application.

Similar to our previous works \cite{chuan}, we consider the same signal model in \cite{Wang} and ignoring clutter. Under assumptions as follows: a MSRS constituted by \(J\) monostatic radars, there are \(U\) resolution cells in the surveillance region, the radar cross section (RCS) of the target is unfluctuating, the background noise is white Gaussian noise, and each radar only transmits one pulse in one observation. For the \textbf{cooperative mode}, as all the received returning signals are jointly processed, \(P_d\) of target locates in the \(u\)th resolution cell (whose position is \((x_u,y_u)\)) under this mode can be calculated as \cite{Shnidman,Cui}:
\begin{align}\label{eq: marcumC}
P_d(\mathbf{\Theta})&=Q_{J\times J}(\sqrt{2 \times \chi_{u}(\mathbf{\Theta})},\sqrt{2 \times \gamma_{T}}),\\
\nonumber
P_{fa}&=e^{-\gamma_{T}}\sum\nolimits_{i=1}^{J\times J-1} \gamma_T^i / i!,
\end{align}
where \(\chi_{u}(\mathbf{\Theta})\) denotes the RTSN of the returning signals that determines the performance of MSRS in cooperative mode, \(Q\) denotes the Marcum Q function, \(\gamma_T\) is the detection threshold which is set according to false alarm probability (\(P_{fa}\)).

For the \textbf{non-cooperative mode}, all the radar nodes of MSRS work independently. The receiver of each radar node only receives the returning signal that transmitted by its own transmitter. Instead of jointly processing all the received signals, each radar node only processes the returning signal received by itself. All the radar nodes have neither cooperation nor communication to each other. Specifically, in this mode, all radar nodes independently detect the target and they all get their own \(P_d\) of target, and thus the \(P_d\) of MSRS is equal to that of the radar node whose \(P_d\) of target is the highest among all radar nodes. Therefore, \(P_d\) of target locates in the \(u\)th resolution cell can be calculated as:
\begin{align}\label{eq: marcumNC}
P_d(\mathbf{\Theta})&=Q(\sqrt{2 \times \chi_{u}(\mathbf{\Theta})},\sqrt{2 \times \gamma_{T}}),\\
\nonumber
P_{fa}&=e^{-\gamma_{T}}\gamma_T,
\end{align}
where \(\chi_{u}(\mathbf{\Theta})\) denotes the RTSN of the returning signals that determines the performance of MSRS in non-cooperative mode. The \(\chi_{u}(\mathbf{\Theta})\) in different mode (i.e., cooperative and non-cooperative mode) are different and we will introduce them as well as their calculation methods in the second part of this subsection (i.e., \textbf{\textit{Objective function of}} \(\mathcal{L_R}\)).

Using the calculation method of \(P_d\) for both the cooperative mode (i.e., use (\ref{eq: marcumC})) and the non-cooperative mode (i.e., use (\ref{eq: marcumNC})), we can calculate the \(P_d\) of every resolution cell within the surveillance region. Assume that we can get \(U^\prime(\mathbf{\Theta})\) effectively covered resolution cells (i.e., in which the \(P_d(\mathbf{\Theta})\)s are higher than \(P_{dt}\)) for a given \(\mathbf{\Theta}\) in the surveillance region. As mentioned before, there are \(U\) resolution cells in surveillance region in total. Then, we can obtain the \(\mathcal{C_R}\) as bellow:
\begin{align}\label{eq: CoverageRatioCac}
\mathcal{C}_\mathcal{R}(\mathbf{\Theta}) & = \frac {\textrm{area}(\mathbb{C}(\mathbf{\Theta}) \cap \mathbb{A})}{\textrm{area}( \mathbb{A})} = \frac {U^\prime(\mathbf{\Theta})}{U}.
\end{align}

\subsubsection{\textbf{Objective function of} \(\mathcal{L_R}\)}\label{sect:snr}
Generally, a MSRS contains many T-R pairs \cite{Haimovich}. For the target in the \(u\)th resolution cell, the T-R pair that transmitted from the transmitter of the \(m\)th radar node and received at the receiver of the \(n\)th radar node, the RTSN of returning signal in this T-R pair is given as:
\begin{equation}\label{eq: standardbio}
\chi^{mn}_u= \frac{P_{tm} \tau A_{tm} A_{rn} \sigma^{mn} \lambda^2 F_{tm}^2 F_{rn}^2}{ {4\pi}^3 K T_s (R_{tm}R_{rn})^2 C_B L_{tm} L_{rn}},
\end{equation}
where \(R_{tm}\), \(R_{rn}\), \(P_{tm}\), \(\tau\), \(A_{tm}\), \(A_{rn}\), \(\sigma^{mn}\), \(\lambda\), \(F_{tm}\), \(F_{rn}\), \(C_B\), \(T_s\), \(K\), \(L_{tm}\), \(L_{rn}\) are respectively transmitter-to-target range, receiver-to-target range, transmitted power of the \(m\)th radar node, pulse width, transmit antenna power gain, receive antenna power gain, bistatic radar target cross section, wavelength, pattern propagation factor for transmitter-to-target path, pattern propagation factor for target-to-receiver path, bandwidth correction factor, receive system noise temperature, Boltzmann's constant, transmit system losses, receive system loss. \(\chi^{mn}_u\) denotes the RTSN of returning signal in this T-R pair.

Generally, most of these above radar parameters are constant and it is unnecessary to take all of them into account. Besides, to concretely set up all these involved parameters is beyond the main scope of this paper. We need to find proper method to calculate the RTSN for each T-R pair conveniently. Similar with our previous work \cite{chuan}, we use a method based on radar range equation to approximatively calculate the RTSN.

\textbf{Firstly}, as we mentioned before, the concerned MSRS is assumed to be constituted by the same kind of monostatic radars. It is reasonable assume that the transmitter and receiver of each radar node share one antenna and all the above radar parameters except transmitted power (we need to optimally allocate them) are the same. Under such assumption, we can get that \(A_{tm}=A_{rn}=A\) and \(L=L_{tm}L_{rn},  \text{for} \hspace{1.5mm} 1 \leqslant n,m \leqslant J\). Furthermore, we ignore the effect of pattern propagation factors (i.e., \(F_{tm} = F_{rn} =1, \text{for} \hspace{1.5mm} 1 \leqslant n,m \leqslant J\)). Therefore, the T-R pair observed at \(n\)th receiver due to the transmission at \(m\)th transmitter is given by\cite{Merrill} (still use \(u\)th resolution cell as example):
\begin{equation}\label{eq: bio}
\chi^{mn}_u =  \frac{P_{tm} \tau A^2 \sigma^{mn} \lambda^2}{ {4\pi}^3 K T_s R_{tm}^2R_{rn}^2 C_B L}.
\end{equation}

\textbf{Then}, it is well known that once the radar system parameters are established, we can get a maximum detection range for each of the concerned radar node. The maximum detection range can be given by  radar range equation \cite{Merrill}:
\begin{equation}\label{eq: alone}
R_{\max} = [ \frac{P_t \tau A^2 \sigma \lambda^2}{ {4\pi}^3 K T_s D_0 C_B L} ]^{\frac{1}{4}},
\end{equation}
where \(\sigma\) is the radar target cross section and \(R_{\max}\) is the maximal detection range when radar node works independently, \(D_0\) is the detectability factor and \(P_t\) is a standard transmitted power satisfied with a fixed performance characterized by \(R_{\max}\), \(\sigma\), \(P_{fa}\) and \(P_{dt}\).

\textbf{Finally}, the RTSN of returning signal in the T-R pair which observed at \(n\)th receiver due to the transmission at \(m\)th transmitter can be obtained by substituting (\ref{eq: alone}) into (\ref{eq: bio}) and with some simple derivation:
\begin{align}\label{eq: snr cac}
\chi^{mn}_u&=D_0 \frac{ \rho_m \sigma_{m,n} (R_{\max})^4}{ \sigma (R_{tm}R_{rn})^2},\\
\rho_m&=\frac{P_{tm}}{P_t}.
\end{align}
Instead of setting specific values, here we allocate the transmitted power by adjusting a ratio (\(\rho\)) between the standard one (\(P_t\)) and the actual transmit ones (\(P_{tm}\)s) of each radar node. For the \textbf{cooperative mode}, there are \(J \times J\) T-R pairs. This is because that each receiver of radar nodes can receive and separate all the returning signals reflected by targets. As jointly processing the received signals from all T-R pairs, the performance of MSRS in this mode is effected by all the T-R pairs. Meanwhile, the noise power of different T-R pairs are the same due to the same kinds of receiver in different radar nodes. Then, the RTSN of the \(u\)th resolution cell can be obtained as:
\begin{align}\label{eq: LSSRC}
\nonumber
\chi_u(\mathbf{\Theta}) &=\frac{E_S^t}{\sigma_0}\\
\nonumber
&=\frac{\sum\nolimits_{m=1}^J \sum\nolimits_{n=1}^J E_S^{mn}}{\sigma_0}\\
\nonumber
&=\sum\nolimits_{m=1}^J \sum\nolimits_{n=1}^J\frac{ E_S^{mn}}{\sigma_0}\\
&=\sum\limits_{m=1}^J \sum\limits_{n=1}^J \chi^{mn}_u,
\end{align}
where \(E_S^t\), \(E_S^{mn}\) and \(\sigma_0\) respectively denote the total signal energy, the signal energy of a T-R pair (i.e., the pair transmitted by \(m\)th node and received by \(n\)th node) and the noise power of each T-R pair.

For the \textbf{non-cooperative mode}, there are \(J\) T-R pairs. This is because that each radar node can only receive and process the signal transmitted by itself. Since all radar nodes works independently without neither cooperation nor communication to each other, the performance of the MSRS is determined by the T-R pair whose own performance is better than that of other T-R pairs. Therefore, we can obtain the \(\chi_r\) as:
\begin{align}\label{eq: LSSRNC}
\chi_u(\mathbf{\Theta}) =\max \limits_{\scriptstyle {m=1,2,\dots,J}\atop\scriptstyle{n=m}} \chi^{mn}_u,
\end{align}
and then we can calculate \(\mathcal{L_R}\) as:
\begin{align}\label{eq: LSFinal}
\mathcal{L_R}(\mathbf{\Theta}) =\min \limits_{u=1,2,\dots,U} \chi_u(\mathbf{\Theta}).
\end{align}
After the introduction of the objective functions, we give the specific form of our problem as below:
\begin{align}\label{Problem Formulation Specific}
&  \hat{\mathbf{\Theta}}= \text{arg} \max \limits_{\mathbf{\Theta}} \mathbf{G} (\mathbf{\Theta}): = (g_1(\mathbf{\Theta}),g_2(\mathbf{\Theta})),\\
\nonumber
&  g_1 (\mathbf{\Theta})=\mathcal{C}_\mathcal{R}(\mathbf{\Theta}) = \frac {\textrm{area}(\mathbb{C}(\mathbf{\Theta}) \cap \mathbb{A})}{\textrm{area}( \mathbb{A})} ,\\
\nonumber
&  g_2 (\mathbf{\Theta})=\mathcal{L}_\mathcal{R}(\mathbf{\Theta}) = \min \limits_{r=1,2,\dots,R}\chi_r(\mathbf{\Theta}),\\
\nonumber
&  s.t. \quad \sum\limits_{j=1}^J P_j=P_T=JP_t.
\end{align}
We remark that the definition of the deployment vector (\textbf{DV}) in this problem is \(\mathbf{\Theta}=[\btheta_1,\dots,\btheta_{j},\dots,\btheta_{J},P_1,\dots,P_{j},\dots,P_J]\), where \(\btheta_{j}=(x_j,y_j)\in \mathbb{B}\) and \(P_j\) denote the positions of antenna and the transmitted power of the \(j\)th radar node, respectively. \(P_T=JP_t\) denotes the fixed total transmitted power of the whole radar system and \(\mathbb{B}\) denotes the placement region that antenna can be placed in. By optimally placing the antenna and allocating the transmitted power of each radar node, the purpose of the optimization problem is to simultaneously enhance the \(\mathcal{C_R}\) and \(\mathcal{L_R}\), with limitation of number and total transmitted power of all radar nodes.

Such problem is not easy to solve: \textbf{Firstly}, since we need to jointly consider the positions of all antennas and the transmitted power of each radar, the problem is of high dimensionality which would cause huge computational load for the optimization processing. \textbf{Secondly}, with potential conflict and differences between the two objectives, the optimization of this problem could be more complicated than conventional single-objective ones. Focusing on these issues, we propose an novel optimization algorithm based MOPSO to solve the proposed problem in the next section.

\section{Introduction of MOPSO and A Novel MOPSO Algorithm}
By utilizing a swarm of candidate solutions (named as particles) to search for optimal solution, PSO has been proved its validity in many applications \cite{Aziz,WangXue,Kulkarni,Kiranyaz,ZhanZhang} since it is originally inspired by social behavior of bird flocking \cite{Kennedy}. Such successes have motivated researchers to extend the usage of PSO into multi-objective optimization problems (MOP) \cite{Coello2002,Parsopoulos,Coello2004,Raquel,Hu,Reyes,Goudos,Chamaani,Bing,Pradhan}. Nevertheless, to our best knowledge, the use of MOPSO for the deployment of MSRS has barely been investigated. Meanwhile, in order to adopt the MOPSO into real applications, some problems still remain to be solved (e.g., proper nearest neighbor density estimation in situation that the values of objective functions are significant different). In this section, some brief introductions of basic theories of MOP and MOPSO will be given first. Then, we will propose a novel MOPSO algorithm for the optimization deployment of MSRS.
\subsection{Introduction of Multi-objective Particle Swarm Optimization}
Obviously, the problem described in (\ref{Problem Formulation Specific}) is a multi-objective optimization problem (MOP). Generally, the objectives (e.g., \(\mathcal{C_R}\) and \(\mathcal{L_R}\)) of MOP are usually in conflict with respect to each other, which means the improvements in some objectives will probably result in degradations of other objectives. Therefore, there is hardly any single ideal optimal solutions for MOP \cite{Hu}. Generally, based on the concept of Pareto-optimal, a group of non-dominated solutions that represent the best possible compromises among the objectives are needed to solve MOP \cite{Coello2002,Parsopoulos,Coello2004,Raquel,Reyes,Hu,Goudos,Chamaani,Bing,Pradhan}. Specifically, we can get a \textbf{OV} (i.e., \(\mathbf{G}(\mathbf{\Theta})\)) once a \textbf{DV} (i.e., \(\mathbf{\Theta}\)) is determined in the context of (\ref{Problem Formulation Specific}), and then the non-dominated solutions are the \textbf{DV}s whose corresponding \textbf{OV}s are non-dominated. In the multi-objective space, the non-dominated \textbf{OV}s constitute a front which is called as Pareto front (denotes by \(\mathcal{P_F}\)). Many techniques (including PSO) are adopted to solve MOPs in both theoretical research \cite{Coello2002,Parsopoulos,Coello2004,Raquel,Reyes,Hu} and real applications \cite{Goudos,Chamaani,Bing,Pradhan}. Under the background of the proposed optimization problem, we will briefly introduce the principle of MOPSO and then proposed a novel MOPSO algorithm.

To solve the proposed optimization problem by MOPSO, each particle stands for a candidate \textbf{DV} and we suppose there are \(S\) particles (means swarm size is \(S\)). Then, the optimization process of MOPSO begins with a random initialization of these \(S\) \textbf{DV}s (i.e., the positions of antennas and transmitted power of transmitters are randomly set up within fixed value ranges). As commonly setting in the MOPSO, an external archive is set up to store the non-dominated solutions (i.e., the non-dominated \textbf{DV}s) \cite{Coello2004}. Then, from the initial \textbf{DV}s, we select the satisfactory ones out and store them in this external archive. Besides, the initial individual best solution of each particle is set to be the initial position of itself, and then the global best solution is selected from the initial external archive with some methods. After the initialization, all particles will update themselves as follow (without losing of generality, we using the \(s\)th particle as example, \( 1 \leq s \leq S \)) in every iteration:
\begin{align}
\nonumber
\bv^s(t+1) =&\omega(t) \times \bv^s(t)+ c_1 \br_1(t) (\bp^s(t)-\boldsymbol{\Theta}^s(t)) \\
             &+ c_2 \br_2(t) (\bp^g(t)-\boldsymbol{\Theta}^s(t))\label{eq: pso position}, \\
\boldsymbol{\Theta}^s(t+1) =& \boldsymbol{\Theta}^s(t) + \bv^s(t+1)\label{eq: pso velocity},
\end{align}
where \(\bv^{s}\), \(\boldsymbol{\Theta}^{s}\) and \(\bp^s\) denote the velocity, the position (i.e., \textbf{DV}) and the individual best solution of \(s\)th particle (i.e., the best \textbf{DV} proposed by the \(s\)th particle so far), \(\bp^g\) denotes the global best solution (i,e, the best \textbf{DV} proposed by all particles so far). \(\omega(t)\)   is inertia weight. \(c_1\) and \(c_2\) are acceleration constants. \(\br_1(t)\) and \(\br_2(t)\) are vectors whose elements are random independent variables in the range [0,1]. The velocity of particle in each dimension is limited by a fixed \(V_{max}\).

After the updating of position and velocity, we calculate the \textbf{OV} of each new particle and then update the individual best solutions as (still using the \(s\)th solution for instance):
\begin{align}
&\bp^s(t+1)=\left\{\begin{array}{ll}
\boldsymbol{\Theta}^s(t+1),&if \hspace{0.5mm} \mathbf{G}(\boldsymbol{\Theta}^s(t+1)) \prec \mathbf{G}(\bp^s(t)),\\
\bp^s(t),  &others.
\end{array}\right.\label{eq: self update}
\end{align}
Other particles will be updated in the same way. Afterwards, similar with \cite{Coello2004}, we update the external archive, and then the global bests (i.e., \(\bp^g\)) will be selected out from the updated external archive with some selection methods. Such iteration process will continue until \(t\) reaches the \(T_{\max}\), where the optimization process will stop and the optimization results (i.e., the \textbf{DV}s that stored in the external archive) will be output.

Generally, since with great importance to affect the search process as well as the optimization performance, the selection of global best solution from the set of non-dominated solutions becomes a key component of the MOPSO approach designment \cite{Reyes}. Many selection methods are proposed in order to get better optimization performance. Within all the published works, the selection method with absolute crowding distance, which is based on the concept of neighbour density estimation, has been proved with its validity in theoretical research \cite{Raquel,Reyes}. Nevertheless, such method may not able to reasonably select out the global best solution when the values of the objective functions are significantly different (e.g., the difference could be several orders of magnitude), which is a common phenomenon in real applications. Focusing on this issue, we proposed a novel MOPSO algorithm with a non-dominated relative crowding distance. Before the introduction of the proposed MOPSO, we would briefly introduce the MOPSO with absolute crowding distance (MOPSO-CD) firstly.

\begin{figure}[t]
\centering
\includegraphics[width=2.4in]{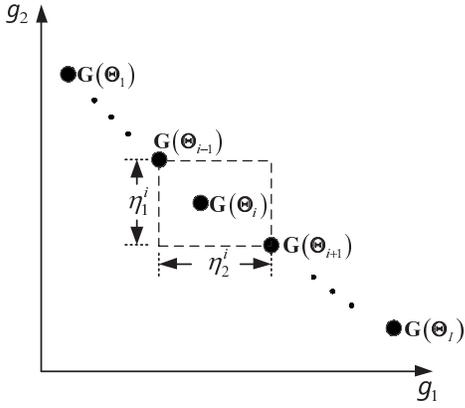}
\caption{An example of the \(i\)th cuboid that belongs to \(\mathbf{G}(\mathbf{\Theta}_i)\) in a two-objective space.}
\label{fig: CD}
\end{figure}

The absolute crowding distance of one solution (i.e., one \textbf{DV}) is the estimation of the size of the largest cuboid (see Fig.\ref{fig: CD}) in multi-objective space. One cuboid can enclose one position point of a specific \textbf{OV} without including any other position points. For instance, the absolute crowding distance of the \(i\)th \textbf{DV} can be calculated as (suppose there are \(I\) non-dominated \textbf{DV}s stored in the external archive):
\begin{align}
\xi_{cd}^i&=\sum\nolimits_{k=1}^{K}\eta_k^i, \quad 2\leqslant i \leqslant I-1\label{eq: acd1},\\
\eta_k^i&=|g_k(\mathbf{\Theta}_{i+1})-g_k(\mathbf{\Theta}_{i-1})|\label{eq: acd2},\\
\xi_{cd}^1&=\xi_{cd}^I=\max\limits_{2\leqslant i \leqslant I-1} \xi_{cd}^i\label{eq: acd3},
\end{align}
where \(\xi_{cd}^i\) denotes the absolute crowding distance of the \(i\)th \textbf{DV} (i.e., \(\mathbf{\Theta}_i\)), \(\eta_k^i\) denotes the sub-crowding-distance in the dimension of \(k\)th objective. The \textbf{DV} with the largest absolute crowding distance has the largest probability to be selected as the global best \cite{Raquel}. As mentioned before, since obtained by directly summing all components, the selection method with absolute crowding distance may face two challenges in real applications:

\textbf{The challenge of value difference:} In real applications, the value of different objective are usually of different orders of magnitude. In (\ref{Problem Formulation Specific}), the value range of \(\mathcal{C_R}\) (i.e., it is in the range of [0,1]) is much smaller than that of \(\mathcal{L_R}\) (e.g., the difference between its lower bound and upper bound could be tens of dB). As obtained by directly summing of all the sub-crowding-distances (i,e,. \(\eta_k^i\)), the absolute crowding distance is mainly effected by the \(\eta_k^i\)s whose values are much larger than the others. Therefore, the crowding distance information of the \(\eta_k^i\)s whose values are small will be covered. Thus, the absolute crowding distance is not adequately enough in such scenarios.

\begin{figure}[t]
\centering
\includegraphics[width=2.4in]{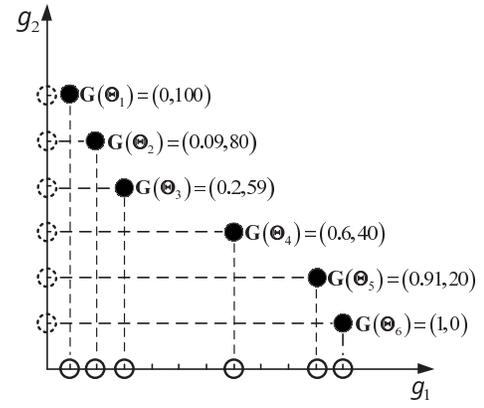}
\caption{An example of 6 position points characterized by 6 \textbf{DV}s in a two-objective space, the two objectives are of different orders of magnitude: the value range of \(g_1\) is [0,1] and the the value range of \(g_2\) is [0,100].}
\label{fig: Cover_Pareto}
\end{figure}

For better understanding, we use an example to illuminate this problem, in the case of maximization problem which has two objectives (\(g_1\) and \(g_2\)), assume that the value range of \(g_1\) is [0,1] and the the value range of \(g_2\) is [0,100]. We demonstrate an \(\mathcal{P_F}\) (which is constituted by 6 non-dominated solutions) and its projections at axis of \(g_1\) and axis of \(g_2\) in Fig.\ref{fig: Cover_Pareto} (we scale down the axis of \(g_2\) for convenient demonstration).

Considering about the value range of each objective and the intervals between neighbouring solutions jointly, we can see that the distribution of the projection on the axis of \(g_2\) is of better uniformity than that of \(g_1\). Specifically, the  \(\mathbf{G(\Theta_4)}\) in \(\mathcal{P_F}\) is obviously sparsest and we should search the space around \(\mathbf{\Theta_4}\) to find new solutions. However, for the aspect of absolute crowding distance, we can see that (in TABLE \uppercase\expandafter{\romannumeral1}) the \(\mathbf{G(\Theta_2)}\) has the biggest absolute crowding distance (ignoring the \(\mathbf{G(\Theta_1)}\) and the \(\mathbf{G(\Theta_6)}\), because they located on the edge of \(\mathcal{P_F}\) and their absolute crowding distance equal to \(\mathbf{G(\Theta_2)}\) according to (\ref{eq: acd3})), which means the space around \(\mathbf{\Theta_2}\) has the highest priority to be searched. Such result of searching priority is unreasonable and it is due to the reason that the value range of \(g_1\) is much smaller than that of \(g_2\) (i.e., at least two orders of magnitude). Therefore, the absolute crowding distance as well as its based selection method of global best solution  are not proper in such kind of real applications.

\begin{table}[hbpt]
\centering
\caption{Absolute Crowding Distance of Solutions in Fig.\ref{fig: Cover_Pareto}}
\begin{tabular}{p{0.5cm}<{\centering}|p{0.8cm}<{\centering}|p{0.8cm}<{\centering}|p{0.8cm}<{\centering}|p{0.8cm}<{\centering}|p{0.8cm}<{\centering}|p{0.8cm}<{\centering}}
\hline
\hline
& \(\mathbf{G(\Theta_1)}\) & \(\mathbf{G(\Theta_2)}\) & \(\mathbf{G(\Theta_3)}\) & \(\mathbf{G(\Theta_4)}\) & \(\mathbf{G(\Theta_5)}\) & \(\mathbf{G(\Theta_6)}\)\\
\hline
\(\xi^i_{cd}\) & 41.2 & 41.2 & 40.51 & 39.71 & 40.4 & 41.2\\
\hline
\end{tabular}
\end{table}

\textbf{The challenge of particle allocation:} Comparing with assigning a same mission to all particles, the cooperation of particles based on labour division is with potential to achieve better optimization performance. For instance, from the principle of MOPSO-CD, since \(\xi_{cd}^1=\xi_{cd}^I=\max\limits_{2\leqslant i \leqslant I-1} \xi_{cd}^i\) and in every iteration that the solution with the largest absolute crowding distance value has the largest probability to be selected as the global best, the particles of MOPSO-CD should search more solutions on the explored \(\mathcal{P_F}\) and extend the value range of each objective function to get a bigger (wider) \(\mathcal{P_F}\) jointly. Obviously, we can divide the optimization goal into two parts (i.e., to find more solutions on \(\mathcal{P_F}\) and to get a bigger \(\mathcal{P_F}\)) and then allocate different swarm of particles to take on these two parts of optimization task.

Addressing on both of these \textbf{two challenges}, we will propose a novel MOPSO algorithm in the next subsection.

\subsection{MOPSO with Non-dominated Relative Crowding Distance}
\textbf{For the challenge of value difference:} we proposed a selection method with non-dominated relative crowding distance. The selection method contains three steps. We will introduce the detail of them in the following.

\textbf{Firstly:} we calculate a crowding distance vector (CDV) and a relative crowding distance for each solution stored in the external archive. For example, for the \(i\)th solution (i.e., \(\mathbf{\Theta}_i\)), the calculation of its CDV is given as follow:
\begin{equation}
\mathbf{H}^i=[\eta_1^i,\eta_2^i,\dots,\eta_K^i]^\text{T},\\
\end{equation}
where \(\mathbf{H}^i\) denotes the CDV of the \(i\)th solution, \(\eta_k^1=2\times|g_k(\mathbf{\Theta}_2)-g_k(\mathbf{\Theta}_1)|\) and \(\eta_k^I=2\times|g_k(\mathbf{\Theta}_I)-g_k(\mathbf{\Theta}_{I-1})|\). The calculation of relative crowding distance is given as follow (still using the \(i\)th solution as example):
\begin{align}
\xi_{rcd}^i=\sum\nolimits_{k=1}^{K}\frac{\eta_k^i}{\max\limits_{1\leqslant i \leqslant I}g_k(\mathbf{\Theta}_{i})-\min\limits_{1\leqslant i \leqslant I}g_k(\mathbf{\Theta}_{i})},
\end{align}
where \(\xi_{rcd}^i\) denotes the relative crowding distance of the \(i\)th solution. \textbf{Then}, we selected out the ones whose CDVs are non-dominated as candidates of global best. Afterwards, within all these candidates, we choose \(Y\) global best solutions whose relative crowding distance are bigger than the rest from these candidates. \textbf{Finally}, we divide the swarm of particles into \(Y\) groups. We allocate one global best solution for each divided group of particles in order to update their positions as well as their velocity. The number of particles contains in each group will be fixed as:
\begin{equation}\label{eq: np}
N_y=S\times \frac{\xi_{rcd}^y}{\sum\nolimits_{y=1}^{Y}\xi_{rcd}^y},
\end{equation}
where \(\xi_{rcd}^y\) is the relative crowding distance of the \(y\)th global best solution, \(S\) is the number of particles in swarm and \(N_y\) is the number of particles in the \(y\)th group.

We still use the example in Fig.\ref{fig: Cover_Pareto} to illuminate this selection process. \textbf{Firstly}, the CDVs (i.e., \(\mathbf{H}\)), the crowding distance (i.e., \(\xi_{cd}\)) and the relative crowding distance (i.e., \(\xi_{rcd}\)) of all the 6 solutions are given out in TABLE \uppercase\expandafter{\romannumeral2}. \textbf{Then}, with non-dominated CDV, the \(\mathbf{\Theta_2}\), \(\mathbf{\Theta_3}\) and \(\mathbf{\Theta_4}\) are qualified to be the candidates. Thus, the global best solution will be selected out according to their \(\xi_{rcd}\) (i.e., the selected ones will be marked with circle). For instance, since with the largest \(\xi_{rcd}\), the \(\mathbf{\Theta_4}\) has the highest priority to be selected as global best in any case, while the \(\mathbf{\Theta_2}\) is with the lowest priority (only when \(Y=3\)) for its smallest \(\xi_{rcd}\). \textbf{Finally}, we assume there are \(100\) particles in total and we give out the result of particle dividing when \(Y=1,2,3\) (i.e., the numbers of particles in the group assigned to each selected solution are given out after each circle). We notice that the \(\mathbf{\Theta_5}\) is not be selected for any case though its relative crowding distance higher than that of the \(\mathbf{\Theta_2}\), it is because that the CDV of the \(\mathbf{\Theta_5}\) is dominated by the \(\mathbf{\Theta_3}\), which excluding its possibility to be selected as global best. We also give out the pseudo code of this selection method in \textbf{Algorithm 1}.

\begin{table}[hbpt]
\centering
\caption{Absolute Crowding Distance, Crowding Distance Vector and Relative Crowding Distance of Solutions in Fig.\ref{fig: Cover_Pareto}}
\begin{tabular}{p{0.8cm}<{\centering}|p{0.5cm}<{\centering}|p{1.3cm}<{\centering}|p{0.5cm}<{\centering}|p{0.8cm}<{\centering}|p{0.8cm}<{\centering}|p{0.8cm}<{\centering}}
\hline
\hline
 & \(\xi^i_{cd}\) & \(\mathbf{H}^i\) & \(\xi^i_{rcd}\) & \(Y=1\) & \(Y=2\) & \(Y=3\)\\
\hline
\(\mathbf{G(\Theta_1)}\) & 41.2 & \([0.18,40]^{\text{T}}\) & 0.58 &  &  &\\
\hline
\(\mathbf{G(\Theta_2)}\) & 41.2 & \(\mathbf{[0.2,41]^{\text{T}}}\) & 0.61 &  &  &  \(\bigcirc\)(23) \\
\hline
\(\mathbf{G(\Theta_3)}\) & 41.51 & \(\mathbf{[0.51,40]^{\text{T}}}\) & 0.91 &  &  \(\bigcirc\)(45)  &  \(\bigcirc\)(35) \\
\hline
\(\mathbf{G(\Theta_4)}\) & 39.71 & \(\mathbf{[0.71,39]^{\text{T}}}\) & 1.1 & \(\bigcirc\)(100) &  \(\bigcirc\)(55)  &  \(\bigcirc\)(42)  \\
\hline
\(\mathbf{G(\Theta_5)}\) & 40.4 & \([0.4,40]^{\text{T}}\) & 0.8 &  &  &\\
\hline
\(\mathbf{G(\Theta_6)}\) & 41.2 & \([0.18,40]^{\text{T}}\) & 0.58 &  &  &\\
\hline
\hline
\end{tabular}
\end{table}

\begin{algorithm}[htbp]
\caption{{Pseudo code of selection method based on non-dominated relative crowding distance}}
\begin{itemize}
  \item \(Y_{\text{actual}}\) is the number of solutions whose \(\mathbf{H}\)s are non-dominated after check all the solutions that stored in external archive
  \item \(Y_{\text{u}}\) is the fixed maximum number of global best solutions
  \item \(Y\) is the number of global best solutions
\end{itemize}
Calculation the \(\mathbf{H}\) and \(\xi_{ncd}\) for every solution that stored in external archive\;
Select out the solutions whose \(\mathbf{H}\)s are non-dominated\;
\If { $Y_{\text{actual}} \leqslant Y_{\text{u}}$ }{
Choose all the \(Y_{\text{actual}}\) solutions as global best solutions\;
\(Y=Y_{\text{actual}}\)\;
\For { $y=1,2,\ldots,Y$ }{
Use (\ref{eq: np}) to calculate \(N_y\) for the \(y\)th selected global best solution\;
}
}
\If { $Y_{\text{actual}} > Y_{\text{u}}$ }{
Sort the \(Y_{\text{actual}}\) solutions according their relative crowding distance (i.e., \(\xi_{rcd}\)) in descending order, pick the first \(Y_u\) ones (whose relative crowding distance are higher than that of the rest) out as global best solutions\;
\(Y=Y_{\text{u}}\)\;
\For { $y=1,2,\ldots,Y$ }{
Use (\ref{eq: np}) to calculate \(N_y\) for the \(y\)th selected global best solution\;
}
}
\end{algorithm}

\textbf{For the challenge of particle allocation:} we reference to a classic type of MOPSO with multi-swarm structure \cite{Parsopoulos}. \(K+1\) swarms of particles are employed for the searching and different swarms has different missions. We use \(K\) of these swarms, which are called as sub-swarms, to enlarge the shape of \(\mathcal{P_F}\). It is well known that the extremum of each objective function constitutes the edge of the obtained \(\mathcal{P_F}\). Then, we let each sub-swarm do a single-objective PSO by employing one of the objectives as fitness function. By assigning each of the concerned objective a swarm of particles to find a better objective function value, the shape of \(\mathcal{P_F}\) could be enlarged effectively. For the rest one swarm, which is called as main swarm, we assign it to find non-dominated solutions within the edge of the obtained \(\mathcal{P_F}\). Thus, with the particle cooperation based on labour dividing, the search process could be of better pertinence and we can get better optimization performance.
\begin{figure}[t]
\centering
\includegraphics[width=3.5in]{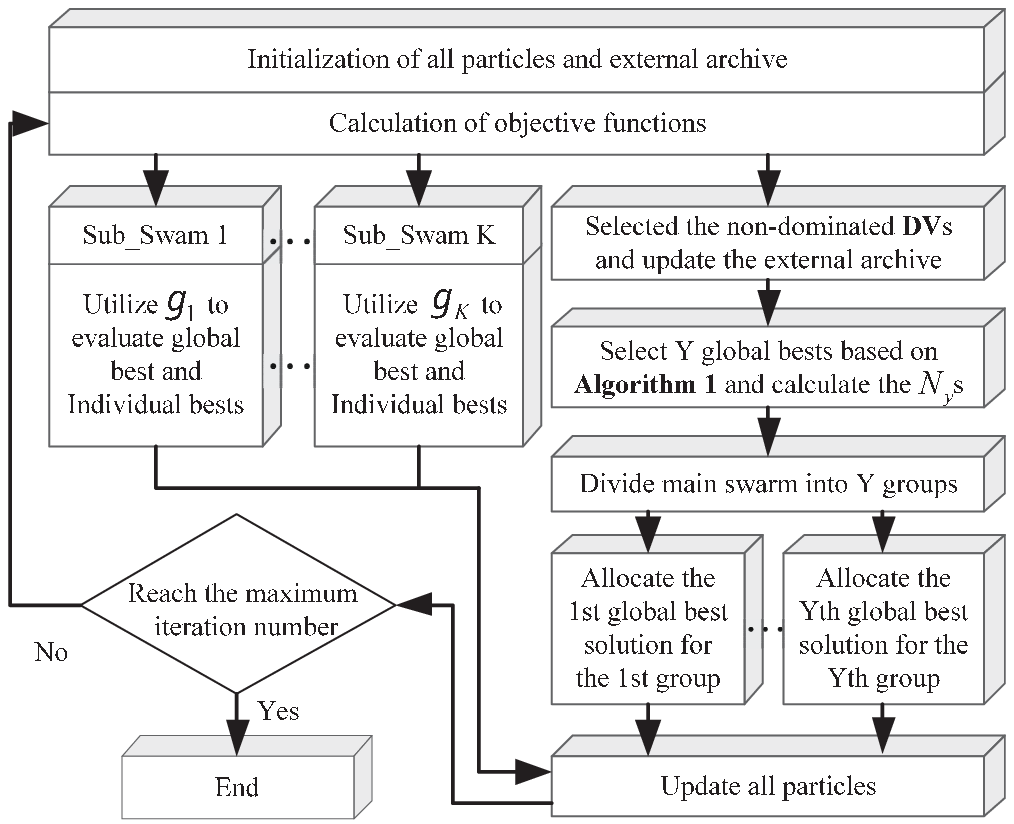}
\caption{Flow diagram of MOPSO-NRCD.}
\label{fig: Flow}
\end{figure}

With using a global best selection method, which is based on a non-dominated relationship of CDV and a relative crowding distance, we call the proposed algorithm as MOPSO with non-dominated relative crowding distance (MOPSO-NRCD) and we give its flow diagram in Fig.\ref{fig: Flow}. We can see that the main swarm and all the sub-swarms work in a parallel way. In every iteration, the main swarm is dynamically divided into \(Y\) groups to chase the \(Y\) selected global best solutions in order to find more non-dominated solutions. Meanwhile, each sub-swarm aims to find better values for its assigned objective function and all sub-swarms work together in order to get a bigger \(\mathcal{P_F}\). The pseudo code of MOPSO-NRCD is given at \textbf{Algorithm 2} (the calculation method of objective functions is given in \textbf{Algorithm 3}, whose detail will be given in the next section).
\begin{algorithm}[htbp]
\nonumber
\caption{{Pseudo code of MOPSO-NRCD}}
\begin{itemize}
  \item \(K\) is the number of the objective functions and the number of sub-swarms
  \item \(S_m\) is the number of particles in main swarm
  \item \(S_k\) is the number of particles in the \(k\)th sub-swarm
\end{itemize}
\For{ $s=1,2,\ldots, S_m$ }{
Initialize particle and its velocity\;
Use \textbf{Algorithm 3} to calculate objective functions of particle\;
Set the initial individual best solution of particle as its initial position\;
}
\For{ $k=1,2,\ldots, K$ }{
\For{ $s=1,2,\dots, S_k$ }{
Initialize particle and its velocity\;
Use \textbf{Algorithm 3} to calculate objective functions of particle\;
Set the initial individual best solution of particle as its initial position\;
}
Use \(g_k\) as fitness function to Initialize the global best solution of the \(k\)th sub-swarm\;
}
Select non-dominated ones from all \textbf{DV} that proposed by particles in main swarm and sub-swarms\;
Store the non-dominated \textbf{DV}s in external archive\;
\For{ $t=1,2,\ldots, T_{max}$}{
Use \textbf{Algorithm 1} to select \(Y\) global bests solutions and divide the main swarm into \(Y\) groups (the \(y\)th group contains \(N_y\) particles)\;
\For{ $y=1,2,\ldots, Y$ }{
utilize the \(y\)th global best solution to update the \(N_y\) particles in \(y\)th group\;
Use \textbf{Algorithm 3} to calculate the objective functions of particle\;
Update the individual best solution of each particle in \(y\)th group\;
}
\For { $k=1,2,\ldots, K$ }{
\For { $s=1,2,\dots, S_k$ }{
Update the velocity and position of particle\;
Use \textbf{Algorithm 3} to calculate the objective functions of particle\;
Use \(g_k\) as fitness function to update the individual best solution of  particle\;
}
Use \(g_k\) as fitness function to update the global best solution of the \(k\)th sub-swarm\;
}
Use the obtained non-dominated \textbf{DV}s in current iteration to update external archive\;
}
Output the solutions that stored in external archive.\
\end{algorithm}

\section{Numerical Results}\label{sect:numerical result}
\begin{table}[hbpt]
\centering
\caption{\label{ParaSet}Parameter Setting of Simulation}
\begin{tabular}{|c | c | c | c|}
\hline
Parameter Name & Value & Parameter Name & Value\\
\hline
\(P_{dt}\) & \(0.8\) & \(P_{fa}\) & \(10^{-6}\)\\
\hline
\(D_0\) & 12.5dB & \(\sigma\) & \(1\)\\
\hline
\(T_{max}\)  & \(2000\) & \(R_{max}\)  & \(6 \hspace{0.5mm} \text{km}\) \\
\hline
\(Y_u\) & 3 & Size of grid & \(2.5 \hspace{0.5mm} \text{km} ^2\)\\
\hline
\(c_1\) & \(2\) & \(c_2\) & \(2\)\\
\hline
\(V_{max}\) & \(4\) & \(S\) of MOPSO-CD & \(200\)\\
\hline
\(S_m\) of MOPSO-NRCD & \(100\) & \(S_k\) of MOPSO-NRCD & \(50\)\\
\hline
\end{tabular}
\end{table}

Before the introduction of simulation results, we firstly give out some common simulation settings in TABLE \ref{ParaSet}. For fairly comparison, the number of the employed particles of both MOPSO-CD and MOPSO-NRCD are set as the same (i.e., the number is set as \(200\)). According to the designed performance and the assumption that every radar transmits one pulse during an observation time, we can find out \(D_0=12.5\) dB \cite{Richards}. We also set that \(\sigma=1 , \sigma_{m,n}=|\alpha|^2\), where \(\alpha\) is assumed to be zero-mean complex gaussian. The \(\omega(t)\) is set to be decreased with iteration time as \(\omega(t)=0.9-0.5*(t/T_{\max})\). Let placement region and surveillance region are coincident (\(\mathbb{A}=\mathbb{B}=50 \hspace{0.5mm} \text{km} \times 50 \hspace{0.5mm} \text{km}\)). The calculation method of the objective functions of (\ref{Problem Formulation Specific}) is given in \textbf{Algorithm 3}.
\begin{algorithm}[htbp]
\caption{{Pseudo code of calculation methods of \(\mathcal{C_R}\) and \(\mathcal{L_R}\) in (\ref{Problem Formulation Specific})}}
\begin{itemize}
  \item \(R\) is the number of resolution cells in surveillance region
  \item \(J\) is the number of radar nodes in MSRS
\end{itemize}
\For { $r=1,2,\ldots,R$ }{
Calculate the transmitter-to-target ranges (i.e., \(R_{tm}\)s, \(m=1,2,\dots,J\)) and the receiver-to-target ranges (i.e., \(R_{rn}\)s, \(n=1,2,\dots,J\)) for each radar nodes\;
Calculate the ratio of signal energy to noise power for every T-R pair by substituting the values of \(D_0\), \(\sigma\). \(\sigma_{m,n}\), and \(R_{max}\) into (\ref{eq: snr cac})\;
Calculate the \(\chi_r\) for one resolution cell with (\ref{eq: LSSRC}) and (\ref{eq: LSSRNC}) for cooperative and non-cooperative mode, respectively\;
Calculate the \(P_d\) by submitting the obtained \(\chi_r\) and the fixed \(P_{fa}\) into (\ref{eq: marcumC}) and (\ref{eq: marcumNC}) for cooperative and non-cooperative mode, respectively\;
}
Calculate the \(\mathcal{C_R}\) and \(\mathcal{L_R}\) by (\ref{eq: CoverageRatioCac}) and (\ref{eq: LSFinal}), respectively\;
\end{algorithm}

\begin{figure}[htbp]
\begin{minipage}{0.48\linewidth}
  \centerline{\includegraphics[width=4.8cm]{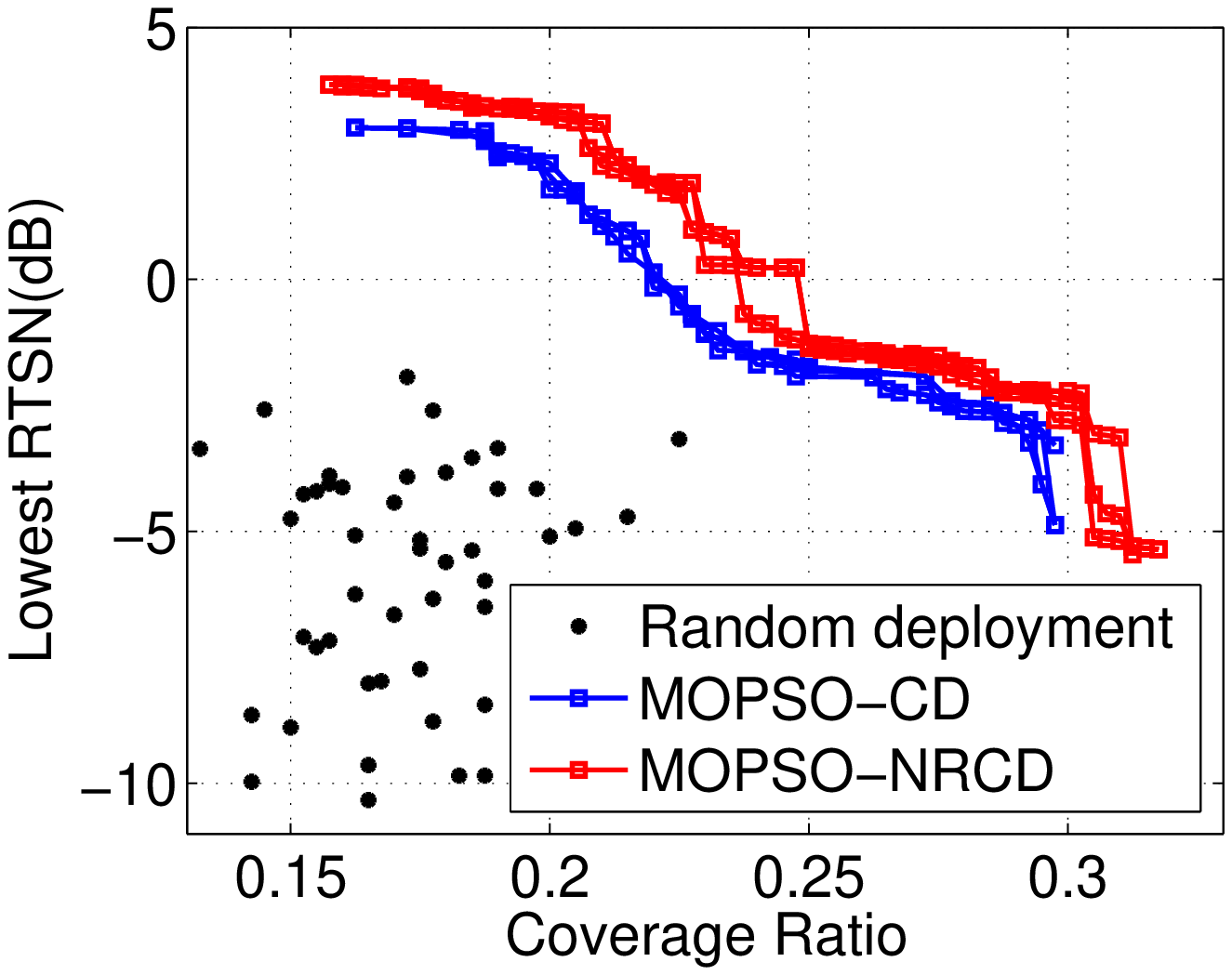}}
  \centerline{(a)}
\end{minipage}
\hfill
\begin{minipage}{0.48\linewidth}
  \centerline{\includegraphics[width=4.8cm]{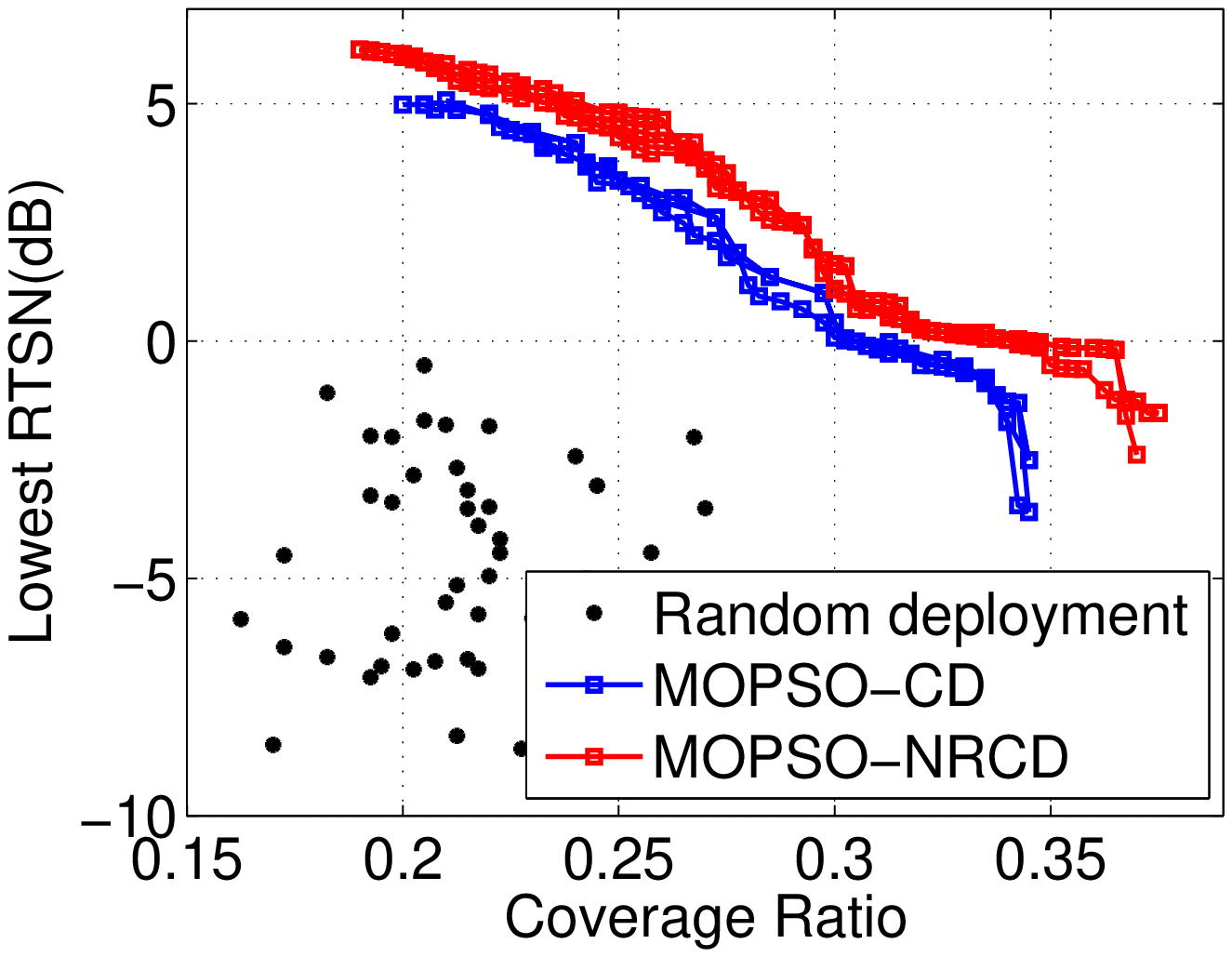}}
  \centerline{(b)}
\end{minipage}
\vfill
\begin{minipage}{0.48\linewidth}
  \centerline{\includegraphics[width=4.8cm]{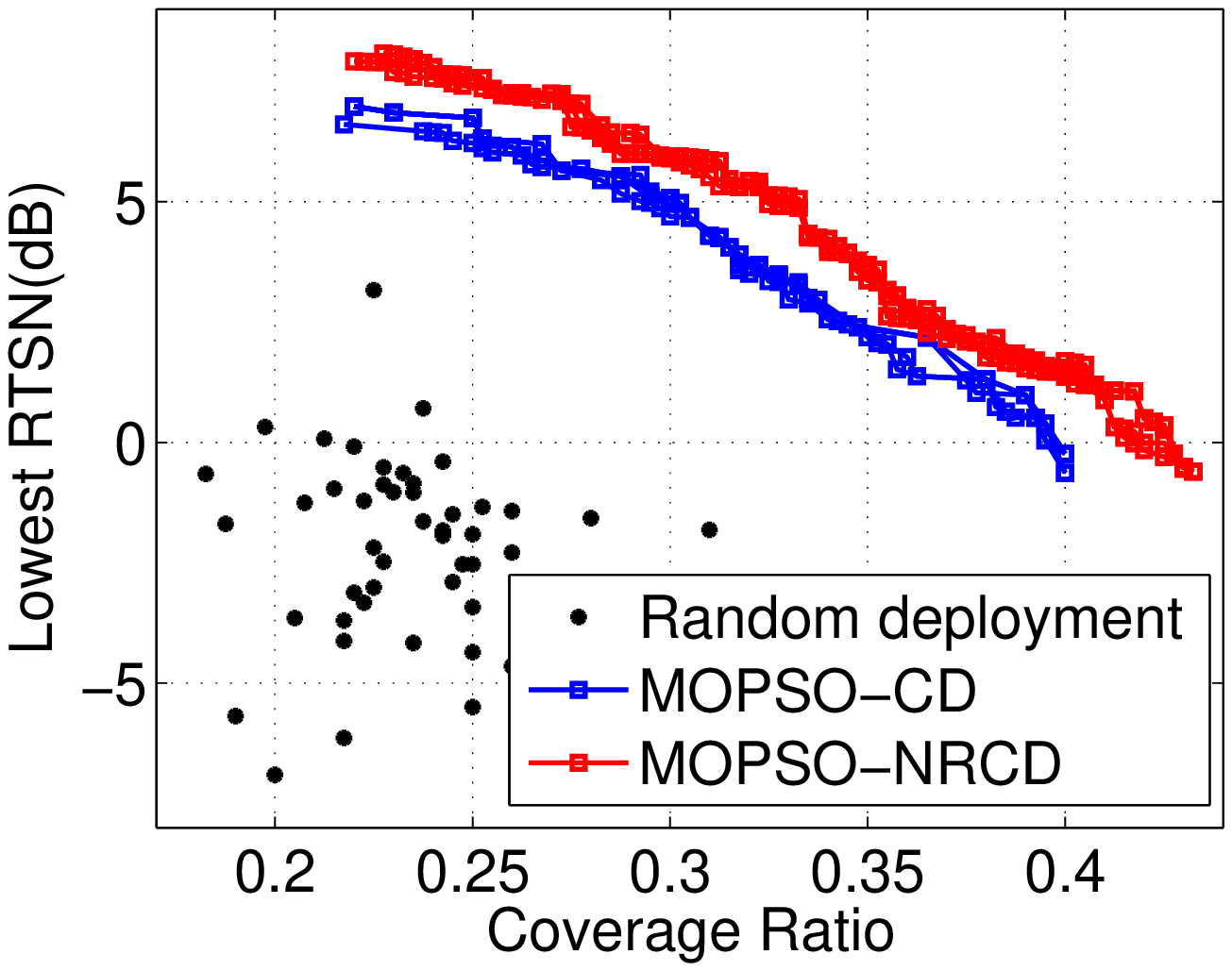}}
  \centerline{(c)}
\end{minipage}
\hfill
\begin{minipage}{0.48\linewidth}
  \centerline{\includegraphics[width=4.8cm]{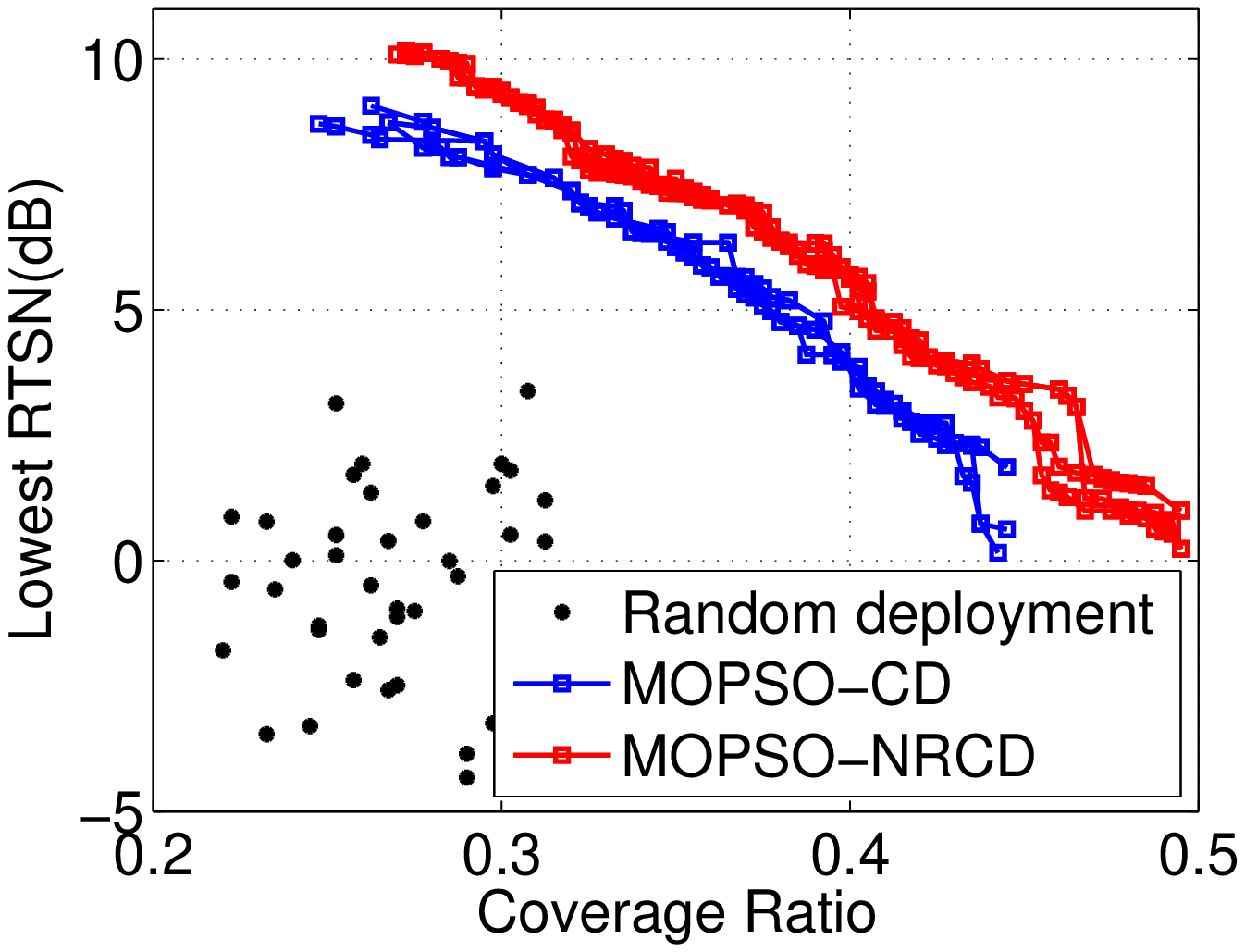}}
  \centerline{(d)}
\end{minipage}
\vfill
\caption{The performance of solutions obtained by random deployment, MOPSO-CD and MOPSO-NRCD under condition of cooperative mode. (a). 5 antennas; (b). 6 antennas; (c). 7 antennas; (d). 8 antennas.}
\label{fig: res1}
\end{figure}

\begin{figure}[htbp]
\begin{minipage}{0.48\linewidth}
  \centerline{\includegraphics[width=4.8cm]{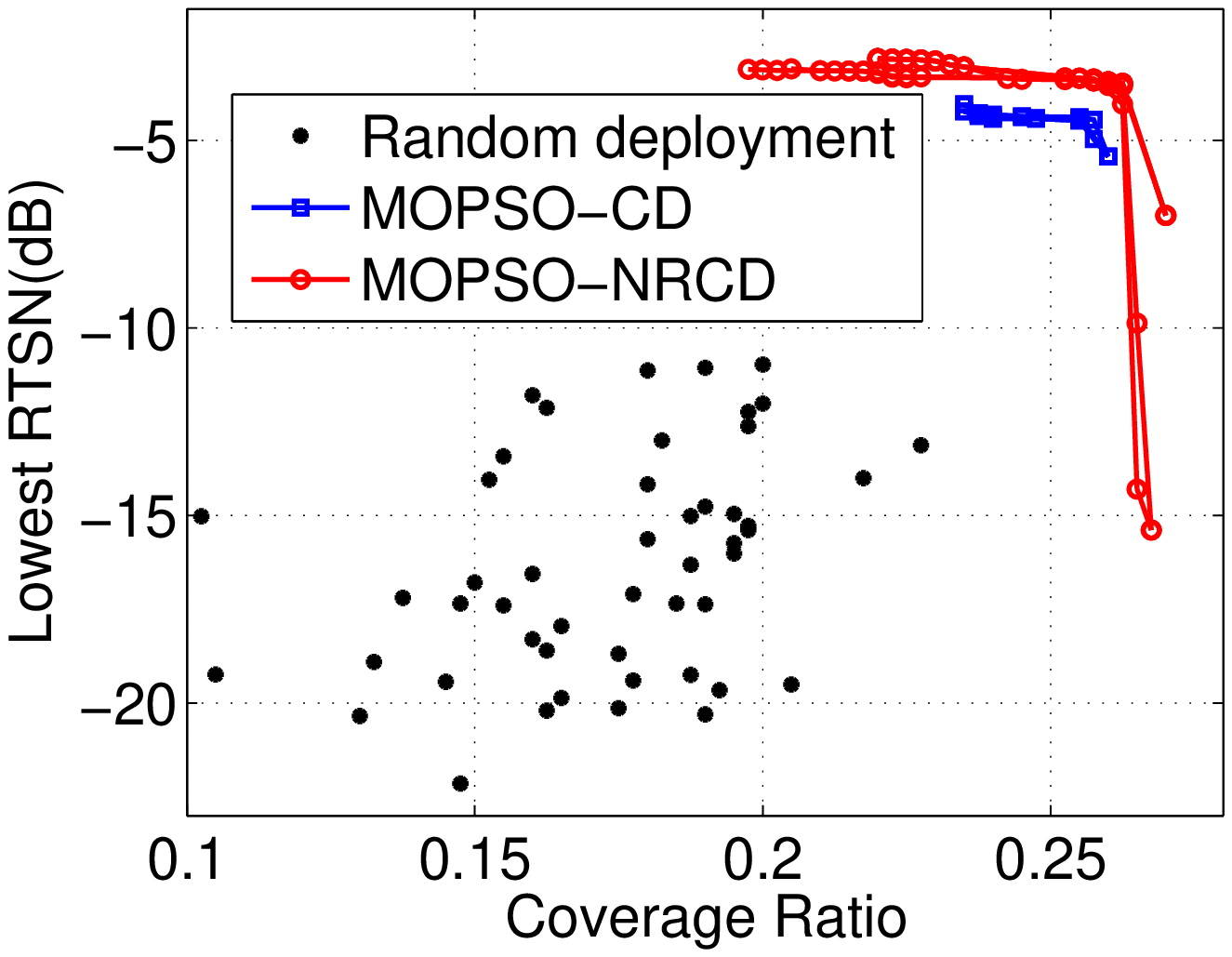}}
  \centerline{(a)}
\end{minipage}
\hfill
\begin{minipage}{0.48\linewidth}
  \centerline{\includegraphics[width=4.8cm]{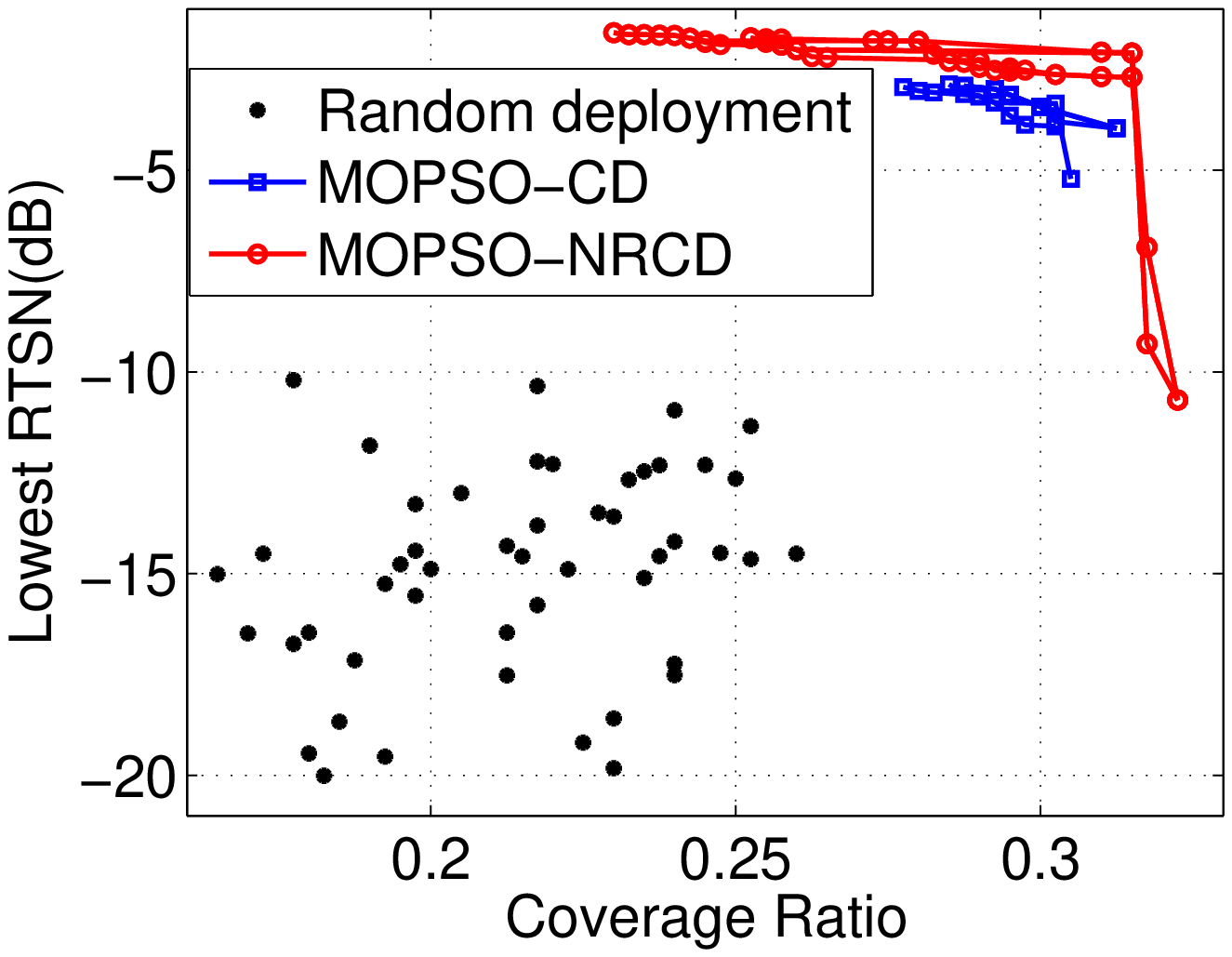}}
  \centerline{(b)}
\end{minipage}
\vfill
\begin{minipage}{0.48\linewidth}
  \centerline{\includegraphics[width=4.8cm]{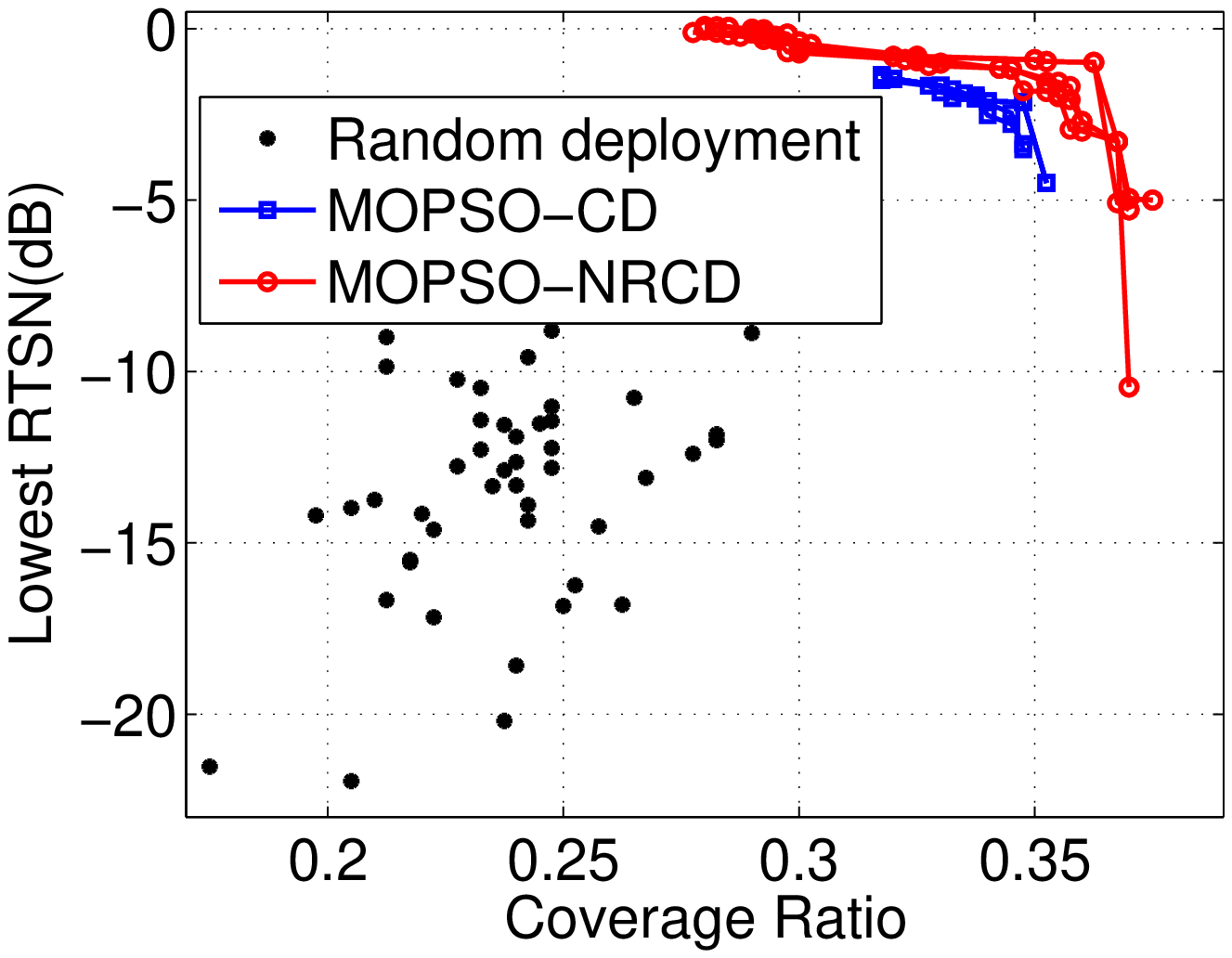}}
  \centerline{(c)}
\end{minipage}
\hfill
\begin{minipage}{0.48\linewidth}
  \centerline{\includegraphics[width=4.8cm]{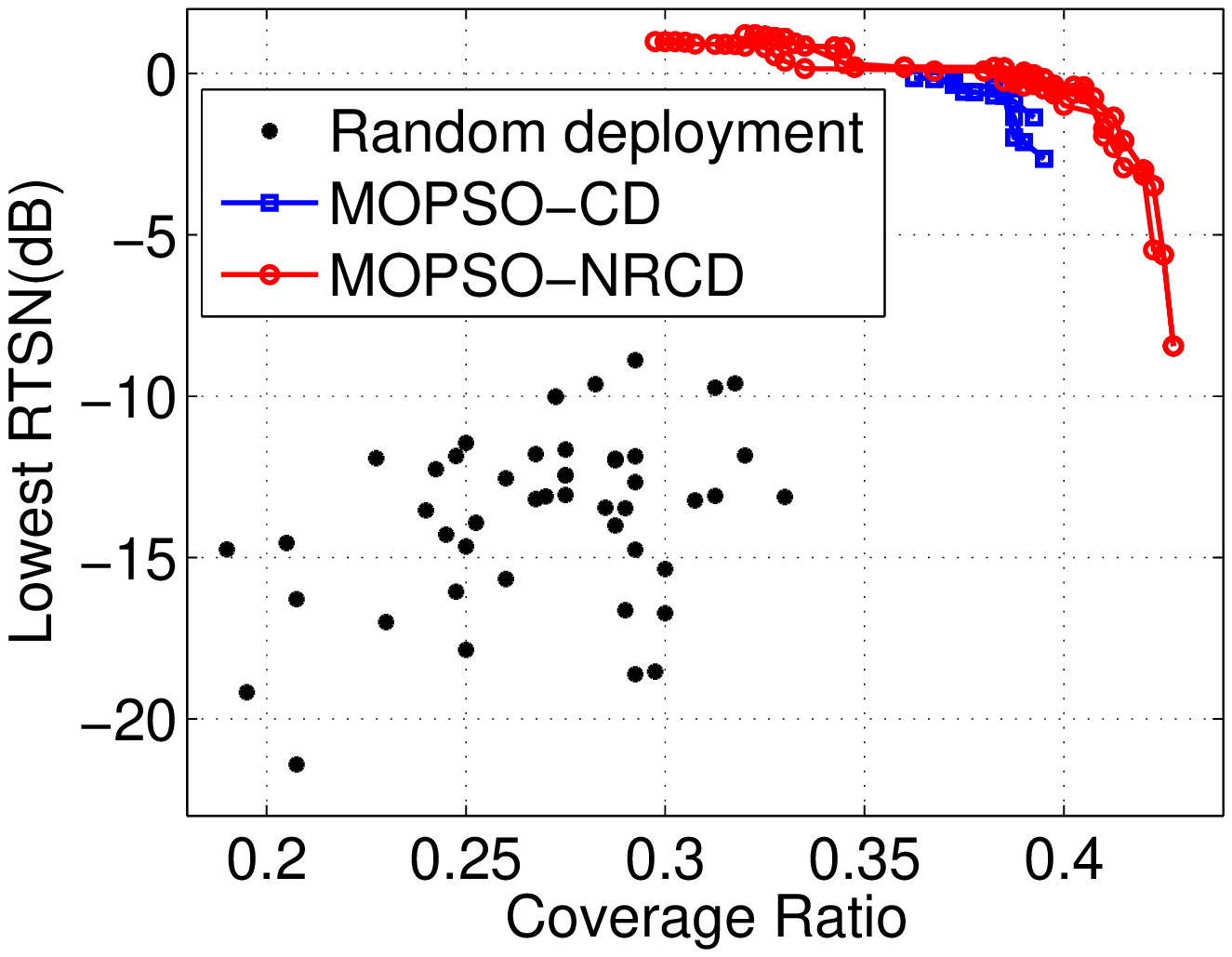}}
  \centerline{(d)}
\end{minipage}
\vfill
\caption{The performance of solutions obtained by random deployment, MOPSO-CD and MOPSO-NRCD under condition of non-cooperative mode. (a). 5 antennas; (b). 6 antennas; (c). 7 antennas; (d). 8 antennas.}
\label{fig: res2}
\end{figure}

We execute 5 independent optimization tests for both MOPSO-CD and NOPSO-NRCD in different simulation cases, which are characterized by different antenna number and different working mode. Meanwhile, we also give the performance of 50 solutions obtained by random deployment in each case for comparison. From Fig. \ref{fig: res1} and Fig. \ref{fig: res2}, we can see that the MOPSO-CD can obtain better solutions comparing with that of random deployment. Specifically, all the solutions of random deployment are dominated by at least one solution obtained by the MOPSO-CD (i.e., the \(\mathcal{C_R}\) and the \(\mathcal{L_R}\) of each solution obtained by random deployment are both smaller than that of at least one solutions obtained by MOPSO-CD). Meanwhile, we can see that each solution obtained by MOPSO-CD are dominated by at least one solution of MOPSO-NRCD in any cases. This proves that the MOPSO-NRCD is capable to achieve better optimization performance than MOPSO-CD.

It is hard to directly quantify the improvement on each concerned objective function (e.g., MOPSO-CD compares with random deployment and MOPSO-NRCD compares with MOPSO-CD) from Fig. \ref{fig: res1} and Fig. \ref{fig: res2}. Therefore, we propose a metric to solve this problem. The introduction of the metric begins with some definition. For comparison, we call the group of solutions with better performance as improved group of solutions (IGS), while the group of solutions with worse performance as control group of solutions (CGS). Assume that there are \(I^{z}\) solutions in IGS that dominate the \(z\)th solution in CGS (assume that CGS contains \(Z\) non-dominated solutions). Then, we can give the calculation method of the proposed metric as:
\begin{equation}
\mathcal{A_I}^k =\frac{1}{Z}\sum\nolimits_{z=1}^{Z}(\frac{1}{I^{z}}\sum\nolimits_{i^{\prime}=1}^{I^{z}}(g_k(\mathbf{\Theta}_{{i^{\prime}}})-g_k(\mathbf{\Theta}_{z}))).
\end{equation}
Obviously, the \(\mathcal{A_I}^k\) represents the average improvement in \(k\)th objective function of IGS comparing with CGS without worsening in other objective functions. Using the obtained solutions in Fig. \ref{fig: res1} and Fig. \ref{fig: res2}, we analyse the \(\mathcal{A_I}^k\)s of MOPSO-CD comparing with random deployment (i.e., the solutions of MOPSO-CD constitute the IGS and the solutions of random deployment constitute the CGS) as well as that of MOPSO-NRCD comparing with MOPSO-CD (i.e., the solutions of MOPSO-NRCD constitute the IGS and the solutions of MOPSO-CD constitute the CGS). The analysing results are shown in TABLE \ref{Fig1} and TABLE \ref{Fig2}. We can see that the optimal solutions of MOPSO-CD can bring significant improvement when comparing with the solutions of random deployment, and thus the validity of MOPSO is proved. Besides, since the solutions of MOPSO-NRCD bringing further improvement in both of these two concerned objectives when comparing with that of MOPSO-CD, the advantages of MOPSO-NRCD is also proved.

\begin{table}[hbpt]
\centering
\setlength{\abovecaptionskip}{0.01pt}
\setlength{\belowcaptionskip}{-0.01pt}
\caption{\label{Fig1}\(\mathcal{A_I}^k\) of MOPSO-CD compares with random deployment And MOPSO-NRCD compares with MOPSO-CD in Fig.\ref{fig: res1}}
\begin{tabular}{p{0.5cm}<{\centering}|p{1.5cm}<{\centering}|p{1.5cm}<{\centering}|p{1.5cm}<{\centering}|p{1.5cm}<{\centering}}
\hline
\hline
\multirow{3}{*}{\(J\)} & \multicolumn{2}{c|}{MOPSO-CD compares} & \multicolumn{2}{c}{MOPSO-NRCD compares} \\
 & \multicolumn{2}{c|}{with random deployment} & \multicolumn{2}{c}{with MOPSO-CD} \\
\cline{2-5}
 & \(\mathcal{L_R}\) (dB) & \(\mathcal{C_R}\) (\%) & \(\mathcal{L_R}\) (dB) & \(\mathcal{C_R}\) (\%) \\
\hline
5 & 5.47 & 0.0616 & 0.71 & 0.0097 \\
\hline
6 & 6.09 & 0.0698 & 0.69 & 0.0137 \\
\hline
7 & 5.97 & 0.0849 & 0.70 & 0.0144 \\
\hline
8 & 5.74 & 0.0922 & 0.81 & 0.0174 \\
\hline
\hline
\end{tabular}
\end{table}

\begin{table}[hbpt]
\centering
\setlength{\abovecaptionskip}{0.01pt}
\setlength{\belowcaptionskip}{-0.01pt}
\caption{\label{Fig2}\(\mathcal{A_I}^k\) of MOPSO-CD compares with random deployment And MOPSO-NRCD compares with MOPSO-CD in Fig.\ref{fig: res2}}
\begin{tabular}{p{0.5cm}<{\centering}|p{1.5cm}<{\centering}|p{1.5cm}<{\centering}|p{1.5cm}<{\centering}|p{1.5cm}<{\centering}}
\hline
\hline
\multirow{3}{*}{\(J\)} & \multicolumn{2}{c|}{MOPSO-CD compares} & \multicolumn{2}{c}{MOPSO-NRCD compares} \\
 & \multicolumn{2}{c|}{with random deployment} & \multicolumn{2}{c}{with MOPSO-CD} \\
\cline{2-5}
 & \(\mathcal{L_R}\) (dB) & \(\mathcal{C_R}\)  & \(\mathcal{L_R}\) (dB) & \(\mathcal{C_R}\)  \\
\hline
5 & 11.61 & 0.0722 & 1.03 & 0.0115 \\
\hline
6 & 11.61 & 0.0823 & 1.07 & 0.0144 \\
\hline
7 & 11.06 & 0.0948 & 0.86 & 0.0151 \\
\hline
8 & 12.52 & 0.1083 & 0.59 & 0.0129 \\
\hline
\hline
\end{tabular}
\end{table}

From another aspect, instead of analysing the concerned objective one by one, we employ the size of dominated space to jointly evaluate the optimization performance of MOPSO-CD and MOPSO-NRCD. The size of dominated space has been introduced by Zitzler in \cite{Zitzler}. It is a measure of the amount of objective space weakly dominated by a given \(\mathcal{P_F}\) which is constituted by a set of solutions. Generally, a bigger size of dominated space means a better optimization performance. To calculate the size of dominated space, we set the reference point as \(\mathcal{L_R}=-15 \hspace{0.5mm} \text{dB}\) and \(\mathcal{C_R}=0.15\). In TABLE \ref{DomSpac}, we can see that the average size of dominated space of MOPSO-NRCD is larger than that of MOPSO-CD in any cases, which improves the advantages of MOPSO-NRCD on the other hand.

\begin{table}[hbpt]
\centering
\setlength{\abovecaptionskip}{0.01pt}
\setlength{\belowcaptionskip}{-0.01pt}
\caption{\label{DomSpac}Average Size of Dominated Space of MOPSO-CD and MOPSO-NRCD in Cooperative Mode and Non-cooperative Mode of 5 Independent Optimization Tests}
\begin{tabular}{c | c | c | c | c}
\hline
\hline
\multirow{2}{*}{\(J\)} & \multicolumn{2}{c|}{Cooperative Mode} & \multicolumn{2}{c}{Non-cooperative Mode} \\
\cline{2-5}
 & MOPSO-CD & MOPSO-NRCD & MOPSO-CD & MOPSO-NRCD \\
\hline
5 & 2.25 & 2.62 & 1.21 & 1.4 \\
\hline
6 & 3.51 & 4.13 & 1.93 & 2.23 \\
\hline
7 & 5 & 5.74 & 2.73 & 3.25 \\
\hline
8 & 6.51 & 7.78 & 3.65 & 4.31 \\
\hline
\hline
\end{tabular}
\end{table}

\section{Conclusion}
An optimization deployment problem of MSRS radar was studied in this paper. The optimization goal is to improve the surveillance performance of a surveillance region by optimally placing the antennas of MSRS and allocating their transmitted power. Firstly, by introducing two objectives, we had formulated the problem under consideration of two typical working mode of MSRS. The corresponding objective functions as well as their calculation methods were also proposed as well. Then, focusing on the challenge of value difference and the challenge of particle allocation, we had proposed a novel MOPSO with non-dominated relative crowding distance (MOPSO-NRCD). Simulation results showed that the proposed algorithm can achieve the better performance than the MOPSO-CD with same number of employed particles for different working mode and different antenna number of MSRS.


\begin{thebibliography}{1}

\bibitem{Griffiths}
H.D. Griffiths, ``Multistatic, MIMO and networked radar: the future of radar sensors?", invited keynote presentation, \textit{EuRAD} conference, Paris, 30 September 2010.

\bibitem{Baker}
H.D. Griffiths and C. Baker, ``Towards the intelligent adaptive radar network," in \textit{Radar Conference (RADAR), 2013 IEEE,} April 2013.

\bibitem{Haimovich}
A. M. Haimovich, R. S. Blum, and L. J. Cimini,  ``MIMO radar with widely separated antennas." \textit{IEEE Signal Process. Mag.}, vol. 25, no. 1, pp. 116-129, Jan. 2008.

\bibitem{QHe}
Q. He, N. H. Lehmann, R. S. Blum, and A. M. Haimovich, ``MIMO radar moving target detection in homogeneous clutter," \textit{IEEE Trans. Aerosp. Electron. Syst.,} vol. 46, no. 3, pp. 1290-1301, Jul. 2010.

\bibitem{Wang}
P. Wang, H. Li, and B. Himed, ``Moving target detection using distributed MIMO radar in clutter with nonhomogeneous power.", \textit{IEEE Trans. Signal Process.}, vol. 59, no. 10, pp. 4809-4820, Oct. 2011.

\bibitem{AMaio}
A. Maio and M. Lops, ``Design principles of MIMO radar detectors,¡± \textit{IEEE Trans. Aerosp. Electron. Syst.,} vol. 43, pp. 886-898, Jul. 2007.

\bibitem{Godrich}
H. Godrich, A. M. Haimovich, and R. S. Blum, ``Target localization accuracy gain in MIMO radar-based systems," \textit{IEEE Trans. Inf. Theory,} vol. 56, pp. 2783-2803, Jun. 2010.

\bibitem{Neng}
L. Neng-Jing, ``Radar ECCMs new area: Anti-stealth and anti-ARM," \textit{IEEE Trans. Aerosp. Electron. Syst.,} vol. 31, no. 3, pp. 1120-1127, Jul. 1995.

\bibitem{He}
Q. He, R. S. Blum, H. Godrich, and A. M. Haimovich, ``Target velocity estimation and antenna placement for MIMO radar with widely separated antennas," \textit{IEEE J. Sel. Topics Signal Process.}, vol. 4, no. 1, pp.79-100, 2010.

\bibitem{Godrich2011}
H. Godrich, A. Petropulu, and H. V. Poor, ``Power allocation strategies for target localization in distributed multiple-radar architectures," \textit{IEEE Trans. Signal Process.,} vol. 59, no. 7, pp. 3226-3240, Jul. 2011.

\bibitem{Chen}
W. Chen and R. M. Narayanan, ``Antenna placement for minimizing target localization error in UWB MIMO noise radar," \textit{IEEE Antennas Wireless Propag. Lett.,} vol. 10, pp. 135-138, 2011.

\bibitem{Chavali}
P. Chavali and A. Nehorai, ``Scheduling and power allocation in a cognitive radar network for multiple-target tracking," \textit{IEEE Trans. Signal Process.,} vol. 60, no. 2, pp. 715-729, Feb. 2012.

\bibitem{Radmard}
M. Radmard, M. M. Chitgarha, M. Nazari-Majd, and M. M. Nayebi, ``Antenna placement and power allocation optimization in MIMO detection." \textit{IEEE Trans. Aerosp. Electron. Syst.}, vol. 50, no. 2, pp. 1468-1478, 2014.

\bibitem{chuan}
Y. Yang, W. Yi, T. Zhang, G. Cui, L. Kong, X. Yang, J. Yang, ``Fast Optimal Antenna Placement for Distributed MIMO Radar with Surveillance Performance," \textit{IEEE Signal Process. Lett.,} vol.22, pp.1955-1959, Nov. 2015

\bibitem{Ke}
L. Ke, Q. Zhang, and R. Battiti, ``MOEA/D-ACO: A multiobjective evolutionary algorithm using decomposition and ant colony,¡± \textit{IEEE Trans. Cybern.,} vol. 43, no. 6, pp. 1845-1859, Dec. 2013.

\bibitem{Jiang}
S. W. Jiang, J. Zhang, Y. S. Ong, A. N. S. Zhang, and P. S. Tan, ``A Simple and Fast Hypervolume Indicator-based Multiobjective
Evolutionary Algorithm¡±, \textit{IEEE Trans. Cybern.,} vol. 45, no. 10, pp. 2202-2213, 2014.

\bibitem{Shim}
V. A. Shim, K. C. Tan, and H. J. Tang, ``Adaptive memetic computing for evolutionary multi-objective optimization¡±, \textit{IEEE Trans. Cybern.,} vol. 45, no. 4, pp. 610-621, 2014.

\bibitem{Merrill}
M. I. Skolnik, \textit{Radar Handbook}, 2nd ed. New York: McGraw-Hill, Jan. 1990.

\bibitem{Shnidman}
D. A. Shnidman, ``Radar detection probabilities and their calculation," \textit{IEEE Trans. Aerosp. Electron. Syst.,} vol. 31, no. 3, pp. 928-950, July 1995.

\bibitem{Cui}
G. Cui, A. De Maio, and Piezzo, M. ``Performance prediction of the incoherent radar detector for correlated generalized Swerling-chi fluctuating targets." \textit{IEEE Trans. Aerosp. Electron. Syst.}, vol. 49, no. 2, pp. 356-368, 2013.

\bibitem{Aziz}
N. Aziz, A. Mohemmed, and M. Alias, ``A wireless sensor network coverage optimization algorithm based on particle swarm optimization and Voronoi diagram," in \textit{Proc. Int. Conf. Networking, Sensing and Control (ICNSC),} Mar. 2009, pp. 602-607.

\bibitem{WangXue}
X. Wang and S. Wang, ``Hierarchical Deployment Optimization for Wireless Sensor Networks," \textit{IEEE Trans. Mobile Computing,} vol. 10, on. 7, pp. 1028-1041, July. 2011.

\bibitem{Kulkarni}
R. V. Kulkarni and G. K. Venayagamoorthy, ¡°Particle swarm optimization in wireless-sensor networks: A brief survey," \textit{IEEE Trans. Syst., Man, Cybern. C, Appl. Rev.,} vol. 41, no. 2, pp. 262-267, 2011.

\bibitem{Kiranyaz}
S. Kiranyaz, T. Ince, A. Yildirim, and M. Gabbouj, ¡°Fractional particle swarm optimization in multidimensional search space,¡± \textit{IEEE Trans. Syst., Man, Cybern., Part B,} vol. 40, no. 2, pp. 298-319, Apr. 2010.

\bibitem{ZhanZhang}
Z. H. Zhan, J. Zhang, Y. Li, and H. S. H. Chung, ``Adaptive particle swarm optimization,¡± \textit{IEEE Trans. Syst., Man, Cybern. B, Cybern.,} vol. 39, no. 6, pp. 1362-1381, Dec. 2009.

\bibitem{Kennedy}
J. Kennedy and R. Eberhart, ``Particle swarm optimization," in \textit{Proc.IEEE Int. Conf. Neural Netw.}, 27 Nov.¨C1 Dec., 1995, vol. 4, pp. 1942-1948

\bibitem{Richards}
M. A. Richards, \textit{Fundementals of Radar Signal Processing.} New York: McGraw-Hill, 2005.

\bibitem{Chernyak}
Chernyak. V, ``Multisite Radar Systems Composed of MIMO Radars," \textit{IEEE AES Syst. Mag.,} vol. 29, pp. 28-37,
Dec. 2014.

\bibitem{Coello2002}
C. A. Coello Coello and M. S. Lechuga, ``MOPSO: A proposal for multiple objective particle swarm optimization," in \textit{Proc. Congr.
Evol. Comput.,} 2002, pp. 1051-1056.

\bibitem{Parsopoulos}
K. E. Parsopoulos and M. N. Vrahatis, ``Particle swarm optimization method in multiobjective problems," in \textit{Proc. ACM Symp. Applied Computing 2002 (SAC 2002),} 2002, pp. 603-607.

\bibitem{Coello2004}
C. A. Coello Coello, G. T. Pulido, and M. S. Lechuga, ``Handling multiple objectives with particle swarm optimization," \textit{IEEE Trans. Evol. Comput.,} vol. 8, no. 3, pp. 256-279, Jun. 2004.

\bibitem{Raquel}
C. R. Raquel and P. C. Nava, ``An effective use of crowding distance in multiobjective particle swarm optimization," in \textit{Proc. Genetic Evol. Comput.,} 2005, pp. 257-264.

\bibitem{Reyes}
M. Reyes-Sierra and C. A. Coello Coello, ``Multi-objective particle swarm optimizers: A survey of the state-of-the-art," \textit{Int. J. Comput. Intell. Res.,} vol. 2, no. 3, pp. 287-308, 2006.

\bibitem{Hu}
W. Hu and G. G. Yen, ``Adaptive multiobjective particle swarm optimization based on parallel cell coordinate system,¡± \textit{IEEE Trans. Evol. Comput.,} vol. 19, no. 1, pp. 1¨C18, Feb. 2015.

\bibitem{Goudos}
S. K. Goudos, Z. D. Zaharis, D. G. Kampitaki, I. T. Rekanos, and C. S. Hilas, ``Pareto optimal design of dual-band base station antenna arrays using multiobjective particle swarm optimization with fitness sharing," \textit{IEEE Trans. Magnetics.,} vol. 45, no. 3, pp. 1522-1525, Mar. 2009.

\bibitem{Chamaani}
S. Chamaani, S. A. Mirtaheri, and M. S. Abrishamian, ``Improvement of time frequency-domain performance of antipodal Vivaldi antenna using multiobjective particle swarm optimization," \textit{IEEE Trans. Antennas Propag.,} vol. 59, no. 5, pp. 1738-1742, May 2011.

\bibitem{Bing}
B. Xue, M. Zhang, W. Browne, ``Particle swarm optimization for feature selec-tion in classification: a multi-objective approach," \textit{IEEE Trans. Cybern.,} vol. 43, no. 6, pp. 1656-1671, Dec. 2012.

\bibitem{Pradhan}
P. M. Pradhan and G. Panda, ``Connectivity constrained wireless sensor deployment using multiobjective evolutionary algorithms and fuzzy decision making," \textit{Ad Hoc Netw.,} vol. 10, no. 6, pp. 1134-1145, 2012.

\bibitem{Zitzler}
E. Zitzler, ``Evolutionary algorithms for multiobjective optimization: Methods and applications," Doctoral dissertation ETH 13398, Swiss Federal Instit. Technol. (ETH), Zurich, Switzerland, 1999.

\bibitem{Sira}
S. P. Sira, D. Cochran, A. Papandreou-Suppappola, D. Morrell, W. Moran, S. D. Howard, and R. Calderbank, ``Adaptive waveform design for improved detection of low-RCS targets in heavy sea clutter," \textit{IEEE J. Sel. Topics Signal Process.,} vol. 1, no. 1, pp. 56-66, Jun. 2007.

\bibitem{Tonissen}
S. M. Tonissen and R. J. Evans, ``Peformance of dynamic programming techniques for track-before-detect,¡± \textit{IEEE Trans. Aerosp. Electron. Syst.,} vol. 32, no. 4, pp. 1440-1451, Oct. 1996.

\bibitem{WYi}
W. Yi, M. Morelande, L. Kong, and J. Yang. ``A computationally efficient particle filter for multitarget tracking using
an independence approximation." \textit{IEEE Trans. on Signal Processing.,} 61(4):843-856, 2013.

\bibitem{Calderbank}
R. Calderbank, S. Howard, and B. Moran, ``Waveform diversity in radar signal processing," \textit{IEEE Signal Process. Mag.,} vol. 26, no. 1, pp. 32-41, Feb. 2009.

\bibitem{Tian}
T. Zhang, G. Cui, L. Kong, W. Yi, X. Yang, ``Phase-Modulated Waveform Evaluation and Selection Strategy in Compound-Gaussian Clutter," \textit{IEEE Trans. Signal Process.,} vol.61, no.5, pp.1143-1148, Mar, 2013


\end{thebibliography}
\end{document}